\documentclass[fleqn,usenatbib]{mnras}

\usepackage{newtxtext,newtxmath}

\usepackage[T1]{fontenc}

\DeclareRobustCommand{\VAN}[3]{#2}
\let\VANthebibliography\thebibliography
\def\thebibliography{\DeclareRobustCommand{\VAN}[3]{##3}\VANthebibliography}

\usepackage{graphicx}	
\usepackage{amsmath}	
\usepackage{acronym}

\usepackage{comment} 
\usepackage{ulem} 

\newcommand{\be}{\begin{equation}}
\newcommand{\ee}{\end{equation}}
\newcommand{\bea}{\begin{eqnarray}}
\newcommand{\eea}{\end{eqnarray}}
\newcommand{\bel}{\begin{align}}
\newcommand{\eel}{\end{align}}

\def\l{\ell}
\def\lm{\ell m}

\def\i{{\rm i}}

\def\Msun{{\rm M_{\odot}}}
\def\GMc2{{\rm G M_{\odot} c^{-2}}}

\def\L{\mathcal{L}}

\def\Egw{E_\text{GW}}

\def\kt2{\kappa^\text{T}_2}

\def\rhonuc{\rho_\mathrm{nuc}}
\def\rhocgs{\mathrm{g~cm^{-3}}}

\def\rhomax{\rho_\mathrm{max}}

\def\tmerg{t_\mathrm{merg}}
\def\ttmerg{t-\tmerg}
\def\Discmass{M_b^\mathrm{disc}}
\def\mej{M_\mathrm{ej}}
\def\nbary{n_\mathrm{b}}
\def\pm{post-merger}

\usepackage{pifont} 

\newcommand{\thc}{\texttt{WhiskyTHC}~}
\newcommand{\MO}{\texttt{THC\_M1}~}

\usepackage{color}
\definecolor{cyan}{rgb}{0,0.9,0.9}
\definecolor{orange}{rgb}{0.9,0.5,0}
\definecolor{magenta}{rgb}{1,0,1}
\definecolor{purple}{rgb}{0.8,0.4,0.8}
\definecolor{gray}{rgb}{0.8242,0.8242,0.8242}

\acrodef{BNS}[BNS]{binary neutron star}
\acrodef{NR}[NR]{Numerical relativity}
\acrodef{GR}[GR]{general relativity}
\acrodef{EoS}[EOS]{equation of state}
\acrodef{GW}[GW]{gravitational wave}
\acrodef{EM}[EM]{electromagnetic}
\acrodef{MHD}[MHD]{magnetohydrodynamics}
\acrodef{NS}[NS]{neutron star}
\acrodef{BH}[BH]{black hole}
\acrodef{LK}[LK]{leakage}
\acrodef{GRB}[SGRB]{short-gamma-ray burst}
\acrodef{GRLES}[GRLES]{general relativistic Large-Eddy-Simulations}

\title[]{Binary neutron star merger simulations with neutrino
transport and turbulent viscosity: impact of different schemes and 
grid resolution}

\author{Francesco \surname{Zappa}$^{1}$, Sebastiano Bernuzzi$^{1}$, David Radice, Albino Perego}

\author[F. Zappa et al.]{
Francesco Zappa,$^{1}$
Sebastiano Bernuzzi,$^{1}$
David Radice$^{2,3,4}$\thanks{Alfred P.~Sloan Fellow}
and Albino Perego$^{5,6}$
\\
$^{1}$Theoretisch-Physikalisches Institut, Friedrich-Schiller-Universit{\"a}t Jena, 07743, Jena, Germany\\
$^{2}$Institute for Gravitation and the Cosmos, The Pennsylvania State University, University Park, PA 16802, USA\\
$^{3}$Department of Physics, The Pennsylvania State University, University Park, PA 16802, USA\\
$^{4}$Department of Astronomy and Astrophysics, The Pennsylvania State University, University Park, PA 16802, USA\\
$^{5}$ Dipartimento di Fisica, Universit\'{a} di Trento, Via Sommarive 14, 38123 Trento, Italy \\
$^{6}$ INFN-TIFPA,Trento Institute for Fundamental Physics and Applications, via Sommarive 14, I-38123 Trento, Italy \\
}

\date{Accepted XXX. Received YYY; in original form ZZZ}

\pubyear{2022}

\begin{document}
\label{firstpage}
\pagerange{\pageref{firstpage}--\pageref{lastpage}}
\maketitle

\begin{abstract}
We present a systematic numerical relativity study of the impact of different treatment of microphysics and grid resolution in binary neutron star mergers. We consider series of simulations at multiple resolutions comparing hydrodynamics, neutrino leakage scheme, leakage augmented with the M0 scheme and the more consistent M1 transport scheme. Additionally, we consider the impact of a sub-grid scheme for turbulent viscosity.
We find that viscosity helps to stabilise the remnant against gravitational collapse but grid resolution has a larger impact than microphysics on the remnant's stability. 
The gravitational wave (GW) energy correlates with the maximum remnant density, that can be thus inferred from GW observations. M1 simulations shows the emergence of a neutrino trapped gas that locally decreases the temperature a few percent when compared to the other simulation series. This out-of-thermodynamics equilibrium effect does not alter the GW emission at the typical resolutions considered for mergers.
Different microphysics treatments impact significantly mass, geometry and composition of the remnant's disc and ejecta. M1 simulations show systematically larger proton fractions.
The different ejecta compositions reflect into the nucleosynthesis yields, that are robust only if both neutrino emission and absorption are simulated.
Synthetic kilonova light curves calculated by means of spherically-symmetric radiation-hydrodynamics evolutions up to 15 days post-merger are mostly sensitive to ejecta's mass and composition; they can be reliably predicted only including the various ejecta components.
We conclude that advanced microphysics in combination with resolutions higher than current standards appear essential for robust long-term evolutions and astrophysical predictions.
\end{abstract}




\section{Introduction}

The joint observation of the \ac{GW} GW170817 and its associated \ac{EM} 
counterparts gave the first direct evidence that \ac{BNS} mergers are at the 
origin of \ac{GRB} and kilonova transients \citep{Monitor:2017mdv,
Abbott:2018wiz,GBM:2017lvd,Abbott:2017wuw,Arcavi:2017xiz,Coulter:2017wya,
Drout:2017ijr,Evans:2017mmy,Hallinan:2017woc,Kasliwal:2017ngb,
Nicholl:2017ahq,Smartt:2017fuw,Soares-santos:2017lru,Tanvir:2017pws,
Troja:2017nqp,Mooley:2018dlz,Ghirlanda:2018uyx,Ruan:2017bha,Lyman:2018qjg}. 
In particular the kilonova counterpart AT2017gfo is 
commonly interpreted as the 
UV/optical/infrared transient generated by radioactive decays of $r$-process 
elements that form in the mass ejected from the merger and the remnant 
\citep{Chornock:2017sdf,Cowperthwaite:2017dyu,Tanaka:2017qxj,
Utsumi:2017cti,Perego:2017wtu,Villar:2017wcc,Waxman:2017sqv,
Metzger:2018uni,Kawaguchi:2018ptg,Breschi:2021tbm}. 
In these neutron rich outflows, 
successive neutron captures produce heavy neutron-rich but unstable nuclei 
\citep[see, e.g.,][for recent reviews]{Cowan:2019pkx,Perego:2021dpw}. The 
latter decay into stable heavy element nuclei, releasing ${\sim}10^{49-50}
$~erg of nuclear energy. The fraction of energy that thermalises inside the 
ejecta is eventually emitted on a timescale of hours-to-months as the 
expanding material becomes transparent. A detailed \textit{ab-inito} 
calculation of this process is a challenging multi-scale and multi-physics 
problem that involves extreme gravity, relativistic \ac{MHD}, and advanced 
microphysics models for the \ac{NS} matter, including neutrino interactions 
and transport. Despite recent efforts, complete models of the mass ejecta and 
the connection to kilonova observations remain very uncertain.

\ac{NR} simulations represent a fundamental approach for the prediction of astrophysical observables from the merger process and its aftermath \citep[see, e.g.,][for recent reviews on the topic]{Radice:2020ddv,Bernuzzi:2020tgt}. On the one hand, simulations are the only means to calculate \ac{GW} from the merger and \pm{} phase. On the other hand, they crucially allow to identify the different mechanisms for mass ejection together with the kinematical and thermodynamical properties of the unbound material.

Weak interactions and neutrino transport are key ingredients in \ac{NR} simulations. Neutrinos with energies up to tens of MeV are prominently produced after the collisional shock between the \ac{NS} cores and, later, in the hottest regions of the merger remnant and accretion disc \citep[see, e.g.,][]{Eichler:1989ve,Ruffert:1996by,Rosswog:2003rv,Sekiguchi:2011zd,Perego:2014fma,Palenzuela:2015dqa,Foucart:2015gaa,Perego:2019adq,Endrizzi:2019trv}. Neutrinos emission is the dominant process reponsible for the cooling of the remnant. Electron antineutrinos
show the largest peak luminosities, which can reach ${\sim}10^{53}$~erg~s$^{-1}$ rather independently on the binary parameters~\citep{Cusinato:2021zin}. Neutrino-matter interactions determine the composition of the dynamical ejecta primarily via reactions $n+e^+\to p+\bar{\nu}_e$ and $n+ \nu_e \to p+e^-$. 
The resulting leptonization process decreases the neutron content in the 
matter, determining
the outcome of the $r$-process nucleosynthesis and the 
color of the kilonova 
\citep{Metzger:2014ila,Martin:2015hxa,Lippuner:2017bfm}.
Absorption of neutrinos on neutrons affects both the geometry and the mass of the dynamical ejecta, especially at high latitudes \citep{Wanajo:2014wha,Foucart:2016rxm,Perego:2017wtu}. Different transport schemes (see below) determine significant differences even in the averaged dynamical ejecta properties \citep{Nedora:2020qtd}. Neutrino absorption in the remnant disc drives a wind on timescales of hundreds milliseconds \pm{} where lighter nuclei (mass numbers $A\lesssim 130$) are synthesised \citep{Dessart:2008zd,Perego:2014fma,Martin:2015hxa,Fujibayashi:2017xsz}. This wind may contribute to the early blue kilonova although its mass is not sufficient to explain the peak of AT2017gfo. More ab-initio simulations are required to robutstly determine the mass and other properties of neutrino-driven winds \citep{Nedora:2020pak,Fujibayashi:2017xsz}. Neutrinos reprocess matter in the density spiral-wave wind that develops from long-lived remnants; this lanthanide-poor material also contributes to a blue transient \citep{Nedora:2019jhl}. 
On seconds timescales, viscosity and neutrino cooling are key processes in 
the development of disc winds \citep{Fernandez:2014bra,Just:2014fka,
Siegel:2017nub,Fujibayashi:2017puw,Radice:2018xqa,Fernandez:2018kax,
Janiuk:2019rrt,Miller:2019dpt,Fujibayashi:2020qda,Just:2021cls}. 
The latter are poorly explored by \ac{NR} simulations and using advanced 
neutrino transport but they are expected to be the main contribution to 
kilonovae like AT2017gfo 
\citep[e.g.,][]{Radice:2018xqa,Miller:2019dpt,Fujibayashi:2020qda}. 
Neutrinos are also expected to play a role in the (yet uncertain) jet-
launching mechanism for \ac{GRB}. On the one hand, for small enough jet 
opening angles, neutrino-antineutrino pair annihilation can deposit the 
required energy \citep[see, e.g.,][]{Eichler:1989ve,Rosswog:2002rt,
Dessart:2008zd,Zalamea:2010ax,Just:2015dba,Perego:2017fho}. 
On the other hand, neutrino absorption in the funnel above the 
remnant contributes to clean this 
region from baryon pollution \citep{Mosta:2020hlh}. 

Neutrino-matter interactions may also impact the high-density regions of the remnant through out-of-equilibrium effects. For example, a trapped neutrino gas can form in the remnant core descreasing the fluid's pressure \citep{Perego:2019adq}. The analysis of \citet{Perego:2019adq} was performed postprocessing a simulation with \ac{LK}+M0 scheme (see below) and found changes in the pressure at the few percent level. 
Interestingly, a more recent postprocessing analysis showed that the presence of muons in the remnant \ac{NS} could affect the trapped neutrino hierarchy and induce variations in the 
remnant pressure up to $7\%$ \citep{Loffredo:2022prq}.
If neutrino trapping occurs, \citet{Alford:2017rxf} proposed that modified-Urca processes can lead to bulk viscous dissipation and to damping of the remnant density oscillations. Recently, some authors argued that these out-of-equilibrium effects are present in hydrodynamics and \ac{LK} simulations and leave a signature in the \pm{} \ac{GW} signal \citep{Most:2022yhe,Hammond:2022uua}. A trapped neutrino gas is observed in the M1 simulations of \citet{Radice:2021jtw}, but no significant out-of-thermodynamic equilibrium effects on the \pm{} dynamics or \ac{GW} emission were observed. All the simulations employed in these works employ rather low grid resolutions that are known to introduce significant uncertainties in the \pm{} dynamics and the \ac{GW} \citep[e.g.,][]{Breschi:2019srl}. Multi-resolution studies employing a consistent neutrino transport and microphysics appear necessary to assess the impact of out-of-equilibrium effects.

The first \ac{BNS} simulations including neutrino effects employed \ac{LK} schemes in either newtonian gravity \citep{Ruffert:1996by,Rosswog:2003tn} or \ac{GR} \citep{Sekiguchi:2010ep,Sekiguchi:2011zd,Neilsen:2014hha,Galeazzi:2013mia,Radice:2016dwd}. \ac{LK} schemes do not solve for the equation transport of neutrinos, but rather they parametrise the matter cooling rate due to neutrinos with a phenomenological formula based on the optical depth. Neutrino reabsorption can be simulated by coupling a \ac{LK} scheme to a truncated multipolar momentum scheme or to ray-tracing algorithms that evolve free-streaming neutrinos in the optically thin regime \citep{Perego:2014fma,Sekiguchi:2015dma,Foucart:2015vpa,Foucart:2015gaa,Radice:2016dwd,Fujibayashi:2017xsz,Radice:2018pdn,Ardevol-Pulpillo:2018btx,Gizzi:2021ssk}. These schemes should be referred to as \ac{LK}+M0 (or \ac{LK}+M1). They avoid stiff terms in the hydrodynamics equations and thus they are computationally efficient while capturing the main physical aspects. More advanced transport schemes are based on the full solution of the truncated moment formalism \citep{Thorne:1981,Shibata:2011kx}. 
M1 grey schemes for \ac{NR} simulations of \ac{BNS} mergers were developed by 
\citet{Foucart:2016rxm} and more recently refined in \citet{Radice:2021jtw}, 
where the complete source terms are implemented. Compared to \ac{LK} schemes, 
M1 schemes are believed to better model the optically thick regime on time 
scales comparable to the cooling timescale although this has not been 
extensively explored in \ac{NR} simulations yet. The simulation of dynamical 
ejecta with the M1 scheme
shows less neutron-rich material than the one 
calculated with \ac{LK}-based scheme, especially at high latitudes 
\citep{Foucart:2016rxm,Radice:2021jtw}. 
The M1 grey scheme 
has been also compared to a Monte-Carlo scheme on short \pm{} timescales to 
find a few percent agreement on key quantities \citep{Foucart:2020qjb}.

\ac{MHD} instabilities and turbulence are expected to affect the matter flow after merger \citep{Kiuchi:2014hja,Kiuchi:2015sga,Kiuchi:2017zzg,deHaas:2022ytm,Combi:2022nhg}. They can impact the outcome of the merger and provide crucial processes for the \ac{GRB} jet-launching mechanism \citep[e.g.][]{Duez:2004nf,Duez:2008rb,Hotokezaka:2013iia,Ciolfi:2019fie}. 
Global large-scale magnetic stresses, if they develop, can boost mass ejecta \citep{Metzger:2018uni,Siegel:2017jug,Siegel:2017nub,deHaas:2022ytm,Combi:2022nhg}. Currently, significant boosts of the mass fluxes can only be achieved by fine-tuning initial configuration or setting unrealistic strength of the magnetic field \citep{Ciolfi:2020hgg,Mosta:2020hlh}. Indeed, one of the main open issues in the simulations is to achieve adequate grid resolution to resolve the amplification of magnetic fields with realistic strenghts and self-consistently obtain turbulent flow \citep{Kiuchi:2017zzg}. Sub-grid models have been recently proposed to ease these simulations \citep{Radice:2017zta,Shibata:2017xht,Aguilera-Miret:2020dhz}. In particular, \citet{Radice:2020ids} proposed a \ac{GRLES} calibrated on very high-resoluition \ac{GR}-\ac{MHD} resolutions simulations of \ac{BNS} from \citet{Kiuchi:2017zzg}.

In this work we perform the first systematic study of the impact of neutrino schemes on the main observables extracted from \ac{BNS} simulations. We study the evolution of an equal-mass \ac{BNS} with component masses $1.3~\Msun$ and a microphysical \ac{EoS} using hydrodynamics and three different neutrino schemes. We consider a \ac{LK}, a \ac{LK}+M0 (hereafter M0) and a M1 scheme. The M0 simulation series is additionally simulated with the \ac{GRLES} scheme to asses the impact of turbulent viscosity. For each physics prescription, we realise a series of simulations at three different resolutions in order to check convergence and robustness of the results. Our goal is to assess the impact of different microphysics schemes and the role of finite grid resolution on the \ac{GW} and \ac{EM} and neutrino emission, and on nucleosynthesis yields.

The rest of the paper is organised as follows.
In section \S\ref{sec:methods}, we describe
our simulations and the different microphysics schemes
we use, as well as the simulation's setup.
In section \S\ref{sec:dynamics}, we discuss the evolution of the system,
the remnant object and the accretion disc.
In section \S\ref{sec:gws} we consider the GW emission
and the detectability of effects on the remnant's core 
from GW observations.
Section \S\ref{sec:ejecta} is devoted to the study of
the dynamical ejecta mass and composition.
In section \S\ref{sec:nuc_KN} we compare the
nucleosynthesis yields and kilonova emission associated to 
the ejecta from our simulations for different microphysics schemes.
In section \S\ref{sec:nu_luminosity} we examine the variations
in neutrino luminosities and average energies comparing M0 and 
M1 schemes.
We summarise and conclude in section \S\ref{sec:conclusion}.

Throughout the text we use latin letters $a,\,b\dots$ as tensor 
indices, where 0 corresponds to the time index and  $1\dots 3$
are the spatial indices. We furthermore
use Einstein convention for the sum over repeated
indices. 
We express masses in units of solar masses, $\Msun$, and temperature and energy 
in MeV. The other quantities are reported in SI or cgs units.

\section{Methods}\label{sec:methods}

\subsection{Matter model, initial data and evolution methods}\label{ssec:initial_data}

NS matter is modelled using the SLy4-SOR \ac{EoS} (hereafter SLy), a finite-temperature, composition-dependent \ac{EoS} 
based on a Skyrme potential for the nucleonic interaction
\citep{Douchin:2001sv,daSilvaSchneider:2017jpg}.
This \ac{EoS} includes baryons (both free and bound in nuclei), electrons, 
positrons and photons as the relevant degrees of freedom.
The SLy \ac{EoS} predicts a maximum Tolman-Oppenheimer-Volkoff (TOV) gravitational
mass of $M_\mathrm{max}^\mathrm{TOV}\approx 2.05~\Msun$ 
and a radius for a $1.4~\Msun$ \ac{NS} of $R_{1.4} \approx 11.9$ km. 
Both these values are compatible with the observations of extremely 
massive millisecond pulsars \citep{Cromartie:2019kug,Fonseca:2021wxt}, 
with results 
obtained by the NICER 
collaboration \citep{Miller:2019cac,Riley:2019yda},  
and with LIGO-Virgo detections \citep{Abbott:2019ebz}; see also \citet{Breschi:2021tbm} for 
a multimessenger analysis based on \ac{NR} data.

Irrotational initial data in quasi-circular orbit are produced with the 
pseudo-spectral multi-domain code \texttt{Lorene} 
\citep{LoreneCode_adj}.
To construct the initial data we use 
the minimum temperature slice $T = 0.01$ MeV of the \ac{EoS} 
used for the evolution. Neutrino-less beta-equilibrium is initially assumed 
inside the two component \ac{NS}s.

The system is evolved 
using the 3+1 Z4c free evolution scheme for Einstein's 
equations \citep{Bernuzzi:2009ex,Hilditch:2012fp}
coupled with the general relativistic hydrodynamics (GRHD) equations. 
NS matter is modelled as a perfect 
fluid with stress-energy tensor 
\be\label{eq:stressenergy}
T_{ab} = (e + p) u_a u_b + p g_{ab}
\ee
where $e$ and $p$ are the energy density and pressure of the fluid, 
and $u_a$ and $g_{ab}$ are
the four-velocity and the spacetime metric, respectively.
The simulations are performed with the \thc~code 
\citep{Radice:2012cu,Radice:2013hxh,Radice:2015nva,Radice:2013xpa,Radice:2016dwd},
which is built on top of the \texttt{Cactus} 
framework \citep{Goodale:2002a,Schnetter:2007rb}. 
In particular, the spacetime is evolved with the \texttt{CTGamma} 
code \citep{Reisswig:2013sqa} which is part of the 
\texttt{Einstein Toolkit} \citep{Loffler:2011ay,EinsteinToolkit}.
The time evolution is performed with the method of lines, using
fourth-order finite-differencing spatial derivatives for the metric
and the strongly-stability preserving third-order Runge-Kutta 
scheme \citep{Gottlieb:2009a} as the time integrator. 
The timestep is set according to the
Courant-Friedrich-Lewy (CFL) criterion and the CFL factor is 
set to $\alpha_\mathrm{CFL} = 0.15$.
Berger-Oliger conservative adaptive mesh refinement (AMR) 
\citep{Berger:1984zza} with sub-cycling in time and refluxing is 
employed \citep{Berger:1989a,Reisswig:2012nc}, as 
provided by the \texttt{Carpet} module of the 
\texttt{Einstein Toolkit} \citep{Schnetter:2003rb}.  

The simulation domain consists of a cube of side ${\sim}3024$ km,
centred at the centre of mass of the
binary system; only the $z\geq0$ portion of the domain is 
simulated and reflection symmetry about the $xy$-plane is used for $z<0$.
The grid setup consists of 7 refinement levels centred 
on the two \ac{NS}s or in the merger remnant, with the finest level 
covering entirely each star.
In this work we distinguish between low resolution 
(LR), standard resolution (SR) and 
high resolution (HR), for which the minimum spacings in the finest 
refinement level are ${\Delta x}_\mathrm{LR} \approx 247$ m,
${\Delta x}_\mathrm{SR} \approx 185$ m, ${\Delta x}_\mathrm{HR} \approx 123$ m.

In \thc~the proton and neutron number densities $n_p$ and $n_n$ are 
evolved separately according to
\be\label{eq:prot_neutr}
\nabla_a(J^a_{p,n}) = R_{p,n}
\ee
where $J^a_{p,n} \equiv n_{p,n} u^a$ is the four-current 
associated to $n_{p,n}$ and
$R_p = - R_n$ is the net lepton number deposition
rate due to absorption and emission of neutrinos 
and antineutrinos. We denote with $\nbary$ the total baryon
number density, such that $\nbary = n_p + n_n$ while $Y_e$ 
is the electron fraction, defined as the net number density 
of electrons and positrons, normalised to $\nbary$.
Under the assumption 
of charge neutrality, $n_p = Y_e \nbary$.
The expressions for $R_{p,n}$ depend on the particular 
neutrino treatment employed, which will be discussed in the next 
subsection.

\subsection{Neutrino and turbulent viscosity schemes}\label{ssec:physics} 

\begin{table}
	\caption{Weak reactions that are considered in this work. 
		$\nu$ denotes a generic neutrino
		species amongst electron neutrino 
		$\nu_e$, electron antineutrino $\bar{\nu}_e$ or
		heavy flavour neutrinos $\nu_x$. 
		The latter is an effective neutrino species containing
	    muon and tau neutrinos and their antineutrinos lumped together.
		$N$ and $A$ indicate respectively nucleons and 
		generic nuclei.}
	\label{tab:nu_reactions}
	\begin{tabular}{cc}
		\hline\hline
		Reaction & Reference\\
		\hline
		$\nu_e + n \leftrightarrow p + e^- $  & \cite{Bruenn:1985en} \\
		$\bar{\nu}_e + p \leftrightarrow n + e^+ $  & \cite{Bruenn:1985en} \\
		$e^+ + e^- \rightarrow \nu + \bar{\nu} $  & \cite{Ruffert:1996by} \\
		$\gamma + \gamma \rightarrow \nu + \bar{\nu} $  & \cite{Ruffert:1996by} \\
		$\nu + N \rightarrow \nu + N $  & \cite{Ruffert:1996by} \\
		$N + N \rightarrow \nu + \bar{\nu} + N + N$  & \cite{Burrows:2004vq} \\
		$\nu + A \rightarrow \nu + A $  & \cite{Shapiro:1983du} \\
		\hline\hline
	\end{tabular}
\end{table}

Weak interactions and neutrino radiation are simulated with 
three different schemes, namely the \ac{LK} scheme, 
the M0 scheme (which is always coupled with the \ac{LK} scheme), and the M1 
transport scheme. In all schemes, three different neutrino species 
are explicitely modelled: $\nu_e$, $\bar{\nu}_e$, and $\nu_x$, where 
the latter is a collective species describing heavy flavour neutrinos 
and antineutrinos. Moreover, all schemes are grey, i.e. the explicit 
dependence on the neutrino energy is integrated out for all the 
relevant quantities.

The \ac{LK} scheme \citep{Galeazzi:2013mia,Radice:2016dwd} accounts 
for the net emission of neutrinos that are produced as a 
result of weak interactions happening during and after the \ac{NS} collision.
The reactions that are considered in our simulations are 
summarised in Tab.~\ref{tab:nu_reactions}.
Due to the large variety of conditions experienced by matter in 
\ac{BNS} mergers,  
neutrinos that are produced in this process 
can be roughly divided in two components.
A first component gets trapped in the high-density and optically 
thick regions of the \ac{NS} remnant, with the possibility
of diffusing out on the diffusion timescale. 
Such component is close to thermodynamical and weak equilibrium 
with matter.
A second component streams freely from the low-density, 
optically thin regions, with a small probability to 
further interact with the surrounding matter.
The \ac{LK} scheme uses a phenomenological formula to interpolate between 
the diffusion rate and the production rate, where the former (latter) 
is the relevant one in optically thick (thin) conditions. 
The scheme crucially relies on the evaluation of the optical 
depth inside the computational domain.   
The resulting effective rates correspond to neutrinos
leaving the system, carrying away energy and lepton number. 
In particular, the particle emission rates correspond to the 
rates appearing on the right-hand side of Eq.~(\ref{eq:prot_neutr}), 
while the total energy emission rate, $Q$, is included in the
simulations as a source term in the Euler equations 
\be\label{eq:Euler}
\nabla_b T^{ab} = Q u^a \, .
\ee
For technical details on the numerical schemes employed for the 
discretization of Eq.~\eqref{eq:prot_neutr} and \eqref{eq:Euler} 
we refer to \citet{Radice:2018pdn}.
We stress that such a \ac{LK} scheme catches the essential cooling effect 
in \ac{NS} matter provided by the emission of neutrinos.
Moreover, it also affects the matter composition by allowing 
the conversion of neutrons into protons, and viceversa.

However, neutrinos are not explicitly transported and the possible interaction of streaming neutrinos with matter in optically thin condition is neglected.
Additionally, no neutrino trapped component is explicitly modelled 
in it (i.e., neutrino radiation is not included in the stress-energy tensor), since the density of particles and energy of equilibrated 
neutrinos are used only to compute the diffusion rates.
The non-inclusion of a neutrino trapped component in the remnant \ac{NS} excludes the correct modelling of
out-of-equilibrium effects that might manifest due to  
the transition from a neutrino-less beta equilibrium 
to a new equilibrium state with the presence of neutrinos.
Finally, the formation and presence of a 
trapped neutrino gas might change the pressure in the remnant
and therefore potentially have an impact on its stability 
\citep{Perego:2019adq}.

The interaction of the free-streaming neutrino component with matter
in optically thin conditions can be simulated in \thc using the M0 scheme, 
as described in \citet{Radice:2016dwd}.
The M0 scheme accounts for possible re-absorption of 
the emitted neutrinos, as computed by the \ac{LK} scheme, 
and the consequent change in matter's composition (i.e., $Y_e$) 
and temperature.
In our simulations the M0 scheme is implemented on a spherical grid
centred at the centre of the computational grid, with outer 
radius ${\sim}756$ km. 

A more appropriate way to include neutrinos in the simulations 
is the M1 scheme, 
which is an approximated approach to neutrino 
transport that applies to neutrino radiation in all relevant regimes.
The Boltzmann equations describing neutrino transport are first 
cast into a system of 3+1 equations, 
similar to the hydrodynamics equations, using a moment-based approach 
\citep{Thorne:1981,Shibata:2011kx}. These equations are 
integrated over the neutrino energy and evolved
consistently coupled to the matter and spacetime equations. 
In the M1 scheme, the terms that describe neutrino
interactions with matter are included directly in the 
stress-energy tensor of Einstein equations.
In this work we use the module $\MO$ implemented in $\thc$, which was
presented in \citet{Radice:2021jtw}. 
For this scheme it is necessary to introduce a closure, i.e. an
expression for the pressure in terms of the energy and the flux. 
We adopt the approximate analytic \textit{Minerbo closure}.
The latter is exact in the optically thick limit
(matter and radiation in thermodynamic
equilibrium) and in the optically thin limit (radiation streaming at the
speed of light in the direction of the radiation flux) if the system has some symmetries (slab, spherical). 
The two limits are then connected by means of the 
\textit{Eddington} factor as described in \citet{Radice:2021jtw}.
The weak interactions that
we consider in $\MO$ are the
same ones included in the \ac{LK} scheme,
listed in Tab.~\ref{tab:nu_reactions}.
To ensure stable runs with the M1 scheme we make the following choices.
Firstly, we set the relative tolerance 
parameter that is used to solve the implicit timestep in the source 
term to $10^{-10}$. Secondly, we additionally enforce local
thermodynamical equilibrium depending on the equilibration
timescale in a specific cell. In particular, 
if for a given cell the corresponding timestep constains more 
than $X$ e-foldings of the equilibration time,
we assume the neutrinos average energies at equilibrium
for the evolution of the neutrinos number densities.
This prevents failures of the runs and the development 
of spurious features in regions of 
high density and low $Y_e$
in the first few ms after collision. The parameter $X$ has been set 
as $20$ for LR and HR run and as $10$ for the SR run.

For a subset of simulations in which we employ the M0 neutrino 
scheme, we additionally include an effective treatment to simulate 
turbulent viscosity with an implementation based on the 
\ac{GRLES} method.
In particular, we consider the effect of magnetic-induced
viscosity, estimated from high-resolution 
\ac{MHD} simulations in 
full \ac{GR} from \citet{Kiuchi:2017zzg}, as described 
in detail in \citet{Radice:2020ids}. 

\subsection{Simulation sample}\label{ssec:binary_system}

In this work, we choose the \ac{NS} component masses and 
the equation of state in such a way that the merger results in
a remnant \ac{NS} close to \ac{BH} collapse.
We aim at finding possible differences due 
to the microphysics and resolution in the evolution
of such border-line case system.
To accomplish this, we pick \ac{NS} component masses of
$M_1 = M_2 = 1.30~\Msun$, 
and  baryonic masses $M_{1b} = M_{2b}= 1.42~\Msun$. The symmetric
mass ratio of the system is 
$\nu := M_1  M_2/\left(M_1 + M_2\right)^2 = 0.25$.
The initial separation is set to $\sim$45 km.
Thus, the \ac{BNS} system has a total initial gravitational
mass $M \approx 2.60~\Msun$ and initial ADM mass and angular 
momentum
$M_\mathrm{ADM} \approx 2.57~\Msun,\,J_\mathrm{ADM} \approx 6.82~\Msun^2$, respectively.

Our study is based on a total of 15 evolutions of the same initial data.
We consider a pure hydro case, in which only
spacetime and hydrodynamics equations are solved, 
that we label as HY.
Preliminary results about these simulations are presented in Appendix~B of \citet{Breschi:2019srl}.
We simulate the binary evolution including the effect of neutrinos 
using only the \ac{LK} scheme, the \ac{LK} scheme coupled with the M0 scheme, 
and the more advanced M1 scheme. The three different types of 
simulations are labelled as \ac{LK}, M0 and M1, respectively. 
We refer to the simulation in which we employ M0 and viscosity as VM0.
Each model is run at the 3 different resolutions defined in 
\S\ref{ssec:initial_data}, namely LR, SR, and HR. 
We refer to a particular run by indicating first the microphysics and then
the resolution; for instance M0-SR is the run with M0 scheme at standard
resolution.
A complete list of all the simulations is reported in the first
column of Tab.~\ref{tab:disk_ejecta}.
The simulations are performed for a minimum of $31$ ms (M1-HR) 
to a maximum of $155$ ms (LK-LR).
Some runs in which a \ac{BH} forms are affected at later time by the
numerical instability described in \citet{Radice:2021jtw} and thus were not 
continued. Simulation data are analyzed to a safe evolution time reported 
in the second column of Tab. \ref{tab:disk_ejecta}; no spurious effects 
are observed until this time.

\section{Remnant dynamics}\label{sec:dynamics}
The two \ac{NS}s revolve for
about 6 orbits before colliding within ${\sim} 14$ ms from the 
beginning of the simulation. 
The moment of merger is conventionally defined as the peak amplitude 
of the (2,2) GW mode, and we label
it as $\tmerg$. The evolution before this moment is referred as the 
\textit{inspiral-merger} phase,
while the evolution after $\tmerg$ is called \textit{\pm{}}.
After merger a remnant NS forms, which survives for at least 
a few tens of ms. In six of our simulations the remnant \ac{NS}
collapses to a \ac{BH} at $\ttmerg \gtrsim 19.9$ ms.

\subsection{Remnant evolution}\label{ssec:remnant_evo}

\begin{figure*}
	\centering 
        \includegraphics[width=0.98\textwidth]{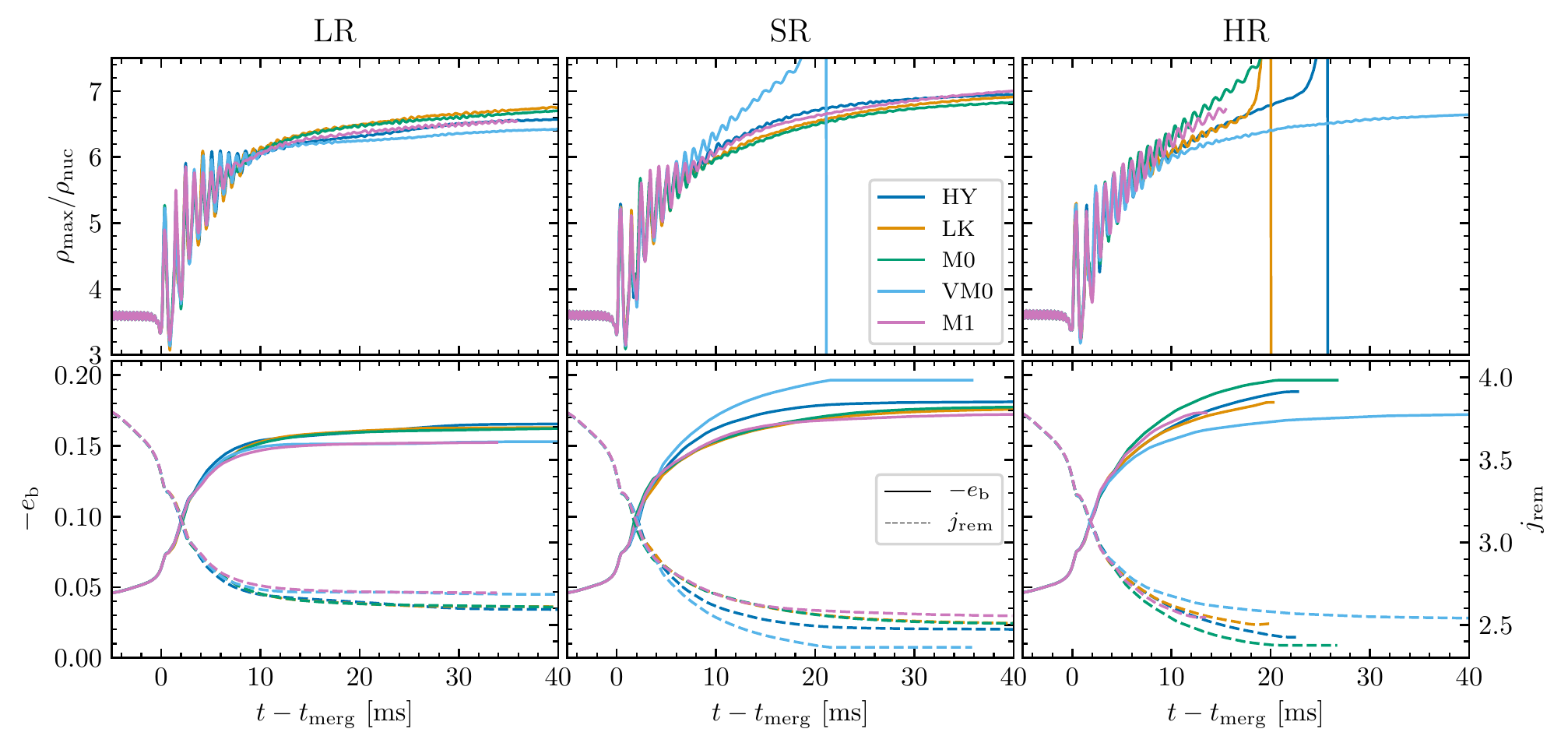}
	\caption{Time evolution of the main quantities describing the 
	dynamics of the system. 
	Top row: maximum rest-mass density in units of the nuclear
	saturation density $\rhonuc \approx 2.3\times 10^{14}~\rhocgs$.
	Bottom row: minus the reduced binding energy of the system (solid lines),
	where $e_b := (M_\mathrm{ADM}-\Egw - M)/(M\nu)$; reduced angular
	momentum of the system 	(dashed lines), where 
	$j_\mathrm{rem} := (J_\mathrm{ADM}-J_\mathrm{GW})/(M^2\nu)$. 
	$M_\mathrm{ADM},\,J_\mathrm{ADM}$ are the ADM mass and angular 
    momentum and $\Egw,\,J_\mathrm{GW}$ are the radiated energy and 
    angular momentum calculated from the multipolar GW 
    \citep{Damour:2011fu,Bernuzzi:2012ci}.
	Time is shifted by the time of merger.}
	\label{fig:recap_merger}
\end{figure*}

The overall remnant evolution is well described
in terms of the maximum rest-mass density, $\rhomax$, (minus) the 
reduced binding energy, $-e_b$, and the reduced angular momentum, 
$j_\mathrm{rem}$, of the system. 
The latter two quantities are defined as
\begin{equation}
e_b := \frac{M_\mathrm{ADM}-\Egw - M}{M\nu}
\end{equation}
and 
\begin{equation}
j_\mathrm{rem} := \frac{J_\mathrm{ADM}-J_\mathrm{GW}}{M^2\nu}
\end{equation}
where $M_\mathrm{ADM},\,J_\mathrm{ADM}$ are the ADM mass 
and angular momentum and $\Egw,\,J_\mathrm{GW}$ 
are the radiated energy and angular momentum 
calculated from the multipolar GW \citep{Damour:2011fu,Bernuzzi:2012ci}.
We report the evolution of these quantities in Fig. \ref{fig:recap_merger},
comparing the different microphysics in each panel and resolution 
effects across the three columns.

For $t < \tmerg$ the evolution is qualitatively and quantitatively
very similar for all the runs. As it can be clearly seen at negative 
times in Fig.~\ref{fig:recap_merger}, the
maximum density (top row), 
reduced binding energy
and angular momentum (bottom row) curves do not 
display any significant differences across the runs.
This is expected, since neutrino production and viscosity effects are
negligible in the two \ac{NS}s.
In this regime, increasing the resolution has the only 
effect of accelerating the merger process and decreasing 
the time of merger, \citep[see, e.g.,][]{Bernuzzi:2011aq,Bernuzzi:2012ci}.  
However, this effect is not visible in Fig. \ref{fig:recap_merger},
because all quantities are shifted by $\tmerg$.

For $t> \tmerg$, $\rhomax$ rapidly increases as the \ac{NS} cores 
merge reaching 
$\gtrsim 6~\rho_\mathrm{nuc}$ within $10$~ms; the damped 
oscillations are caused by the bounces of the two cores in 
the process.
At about $10$ ms \pm{}, the outcome of the GW-dominated (early) \pm{}
phase is a remnant \ac{NS}, 
formed by a core that is slowly rotating surrounded by a
rapidly rotating envelope.
The absolute value of the
binding energy after $\tmerg$ measures the compactness of the 
remnant  \ac{NS} and it increases 
in time due to the emission of gravitational energy. 
The bottom row of Fig. \ref{fig:recap_merger} shows that
most of the emitted gravitational energy and 
angular momentum are radiated within $\ttmerg \approx 10$ ms 
\citep{bernuzzi_2016_57844,Zappa:2017xba}. 
Comparing to the top row, this period coincides with the 
time in which the large oscillations of $\rhomax$ are strongly 
dampened and the remnant \ac{NS} stabilises or collapses. 
The physical explanation is that the 
remnant \ac{NS} has a large and rapidly evolving quadrupole 
momentum and is therefore an efficient
emitter of gravitational radiation.
The emission increases 
the remnant's compactness and 
reduces its angular momentum, thus driving the remnant NS 
towards axisymmetry and eventually stationarity.
Overall, the gravitational energy and angular 
momentum emission show qualitatively a similar evolution for all the 
runs. In all cases about the same values of $-e_\mathrm{b}\approx 0.12$ 
and $j_\mathrm{rem}\approx 2.9$ are reached at $\ttmerg \approx 5$ ms, and
after this time some differences develop among the runs.

\begin{table*}
\caption{Main properties of the remnant disc and ejecta for
all simulations. The end time of the simulation $t_\mathrm{end}$ and
of \ac{BH} collapse $t_\mathrm{BH}$ are measured
with respect to merger. Disc mass $\Discmass$ and ejecta
mass $\mej$ are baryonic masses and are 
expressed in solar masses.
The ejecta quantities are extracted with the Bernoulli criterion
on a spherical surface at $443$ km. 
The electron 
fraction $Y_e$ and the specific entropy $s$ are reported as 
mass-weighted averages. 
The emission angle is calculated as the mass-weighted root 
mean square (RMS) of the emission latitudes.
The analysis is performed until $\ttmerg = 20.3$ ms, 
corresponding to the earliest 
$t_\mathrm{end}$ of our set of simulations, i.e. to the run
LK-HR. The only exception is M1-HR, for which
we use the last available time $t_\mathrm{end} = 15.4$ ms \pm{}.}
\label{tab:disk_ejecta}
\resizebox{1\textwidth}{!}{
\begin{tabular}{ccccccccccc}
\hline\hline
Simulation&$t_\mathrm{end}$ [ms]&$t_\mathrm{BH}$ [ms]&$M^\mathrm{disk}_b[\mathrm{M}_\odot]$& &$M_\mathrm{ej}[\mathrm{M}_\odot]$&$M_\mathrm{ej}^{v \geq 0.6c}[\mathrm{M}_\odot]$&$\left\langle Y_e \right\rangle$&$\theta_\mathrm{ej}^\mathrm{RMS}[{}^\circ]$&$v_\mathrm{ej,~\infty}/c$&$\left\langle s \right\rangle[k_\mathrm{B}/\mathrm{bar}]$\\ 
\hline
\texttt{HY-LR} &109 &- &$1.85\times 10^{-1}$ &&$1.10\times 10^{-2}$ &$1.11\times 10^{-5}$ &0.05 &34 &0.16 &16\\ 
\texttt{LK-LR} &140 &- &$1.76\times 10^{-1}$ &&$2.41\times 10^{-3}$ &$8.60\times 10^{-6}$ &0.13 &28 &0.18 &13\\ 
\texttt{M0-LR} &94 &- &$1.57\times 10^{-1}$ &&$6.70\times 10^{-3}$ &$1.34\times 10^{-5}$ &0.23 &34 &0.16 &17\\ 
\texttt{VM0-LR} &104 &- &$1.80\times 10^{-1}$ &&$6.44\times 10^{-3}$ &$1.48\times 10^{-5}$ &0.23 &34 &0.15 &17\\ 
\texttt{M1-LR} &35.8 &- &$2.42\times 10^{-1}$ &&$6.59\times 10^{-3}$ &$2.02\times 10^{-6}$ &0.24 &36 &0.17 &16\\ 
\hline
\texttt{HY-SR} &109 &- &$1.64\times 10^{-1}$ &&$8.43\times 10^{-3}$ &$2.73\times 10^{-5}$ &0.049 &33 &0.19 &17\\ 
\texttt{LK-SR} &114 &- &$8.14\times 10^{-2}$ &&$2.35\times 10^{-3}$ &$1.23\times 10^{-5}$ &0.16 &30 &0.21 &14\\ 
\texttt{M0-SR} &64.3 &64 &$7.55\times 10^{-2}$ &&$5.85\times 10^{-3}$ &$3.92\times 10^{-5}$ &0.22 &32 &0.18 &16\\ 
\texttt{VM0-SR} &35.8 &21 &$7.58\times 10^{-2}$ &&$4.02\times 10^{-3}$ &$3.09\times 10^{-5}$ &0.23 &33 &0.19 &18\\ 
\texttt{M1-SR} &41.8 &- &$1.51\times 10^{-1}$ &&$4.13\times 10^{-3}$ &$1.29\times 10^{-5}$ &0.24 &37 &0.19 &18\\ 
\hline
\texttt{HY-HR} &27.2 &25.6 &$1.10\times 10^{-1}$ &&$7.20\times 10^{-3}$ &$2.44\times 10^{-5}$ &0.044 &34 &0.19 &18\\ 
\texttt{LK-HR} &20.3 &19.9 &$6.77\times 10^{-2}$ &&$1.92\times 10^{-3}$ &$1.47\times 10^{-6}$ &0.17 &29 &0.2 &16\\ 
\texttt{M0-HR} &28.6 &20.2 &$8.98\times 10^{-2}$ &&$5.11\times 10^{-3}$ &$7.96\times 10^{-6}$ &0.26 &34 &0.16 &18\\ 
\texttt{VM0-HR} &61.3 &60.9 &$9.46\times 10^{-2}$ &&$6.14\times 10^{-3}$ &$2.80\times 10^{-5}$ &0.24 &34 &0.16 &18\\ 
\texttt{M1-HR} &15.4 &- &$9.29\times 10^{-2}$ &&$4.09\times 10^{-3}$ &$4.43\times 10^{-6}$ &0.27 &33 &0.22 &18\\ 
\hline
\hline
\end{tabular}
}

\end{table*}

During the GW-dominated phase, turbulent viscosity has the largest 
impact on the remnant's core dynamics among
all the other microphysics prescriptions.
In particular $\rhomax$ and $-e_b$ in VM0-LR and VM0-HR
runs are comparably smaller with respect to the other runs at
the same resolution, especially at later times. This effect is
due to the fact that viscosity transports angular momentum between the 
slowly rotating core of the remnant
and the rapidly rotating envelope. Consequently, the core 
can acquire angular momentum at the expenses of the 
envelope, gaining more rotational support. 
This effect decreases the central density of the remnant star,
making it more stable \citep{Radice:2017zta,Shibata:2017jyf}.

The grid resolution has a significant impact on the fate of the remnant. 
LR simulations present the smallest GW emission, which leads to a 
less compact and more rotationally supported remnant \ac{NS}.
At LR, gravitational collapse is never observed within the simulated time. 
At higher resolution, we note overall larger 
binding energies and smaller remnant angular momenta for all 
the runs comparing one by one to the LR simulations.
For the M0-SR case \ac{BH} collapse happens at ${\sim} 64$ ms \pm{}
(see third column of Tab. \ref{tab:disk_ejecta}), while  
\ac{BH} formation is not observed for HY-SR, LK-SR and M1-SR suns within the 
end of the simulations.
The HR simulations show the largest absolute values
of binding energies and this determines 
the largest compactnesses for the 
remnant stars. As a consequence, we observe
\ac{BH} formation as early as $\ttmerg \approx 20$ ms for \ac{LK} and M0 runs 
and $\ttmerg \approx 26$ ms for HY case. In VM0-HR the \ac{BH} collapse 
is delayed of about $40$ ms with respect to the other HR runs, due to
the viscosity effects described above. 
The runs that employ M1 transport scheme
show a monotonic behaviour with resolution in $\rhomax$,  
$-e_b$, which increase with resolution, while $j_\mathrm{rem}$ 
decreases.

The run VM0-SR has an unexpected 
behaviour. The remnant \ac{NS} collapses quite early, 
around 20 ms \pm{}. 
Comparing to VM0-LR and VM0-HR runs, 
the density oscillations at $8-10$ ms appear less dampened 
and $\rhomax$ keeps increasing until the \ac{NS} eventually collapses. 
This behaviour has never been observed in 
previous works where viscosity was included in the same way 
\citep{Radice:2020ids,Bernuzzi:2020txg}.
We speculate this result is related to the  
specific simulation setup that, for this particular BNS,
is not yet in a convergent regime at SR. 
Higher resolution simulations would be required 
to explore the possibility of obtaining consistent results.
We leave this investigation to future work.

\begin{figure}
	\centering 
        \includegraphics[width=0.48\textwidth]{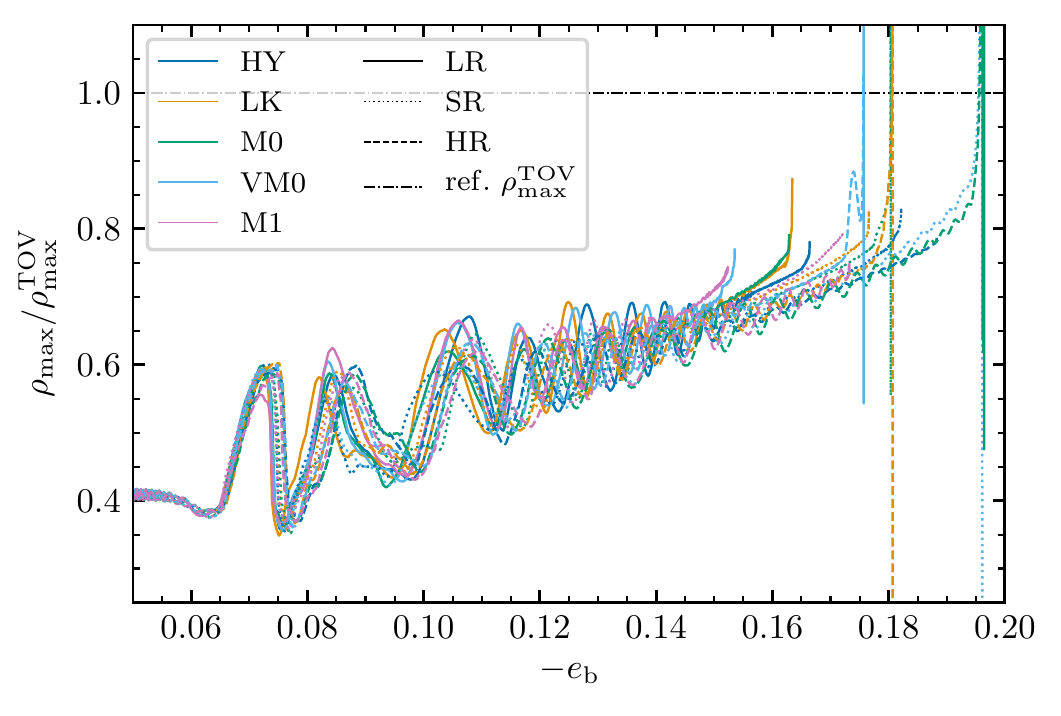}
	\caption{Correlation between the binding energy of the system and
	the maximum rest-mass density. The latter is rescaled by the
	central density of the maximum-mass TOV star predicted by the SLy
	\ac{EoS}.}
	\label{fig:rho_ebind}
\end{figure}

Our results highlight that the analysis of the merger dynamics in terms 
of $\rho$ and energetics is weakly dependent on the particular 
microphysics setup of the simulations and thus it robustly captures 
the merger dynamics. This is summarised 
considering the gauge invariant $\rhomax(-e_b)$ curves in 
Fig.~\ref{fig:rho_ebind}.
The plot shows that the two quantities are clearly
correlated, which implies that $\rhomax$ can be, in 
principle, estimated from a measurement of the 
total \ac{GW} radiated energy \citep{Radice:2016rys}.
The robustness of the correlation showed in Fig. \ref{fig:rho_ebind}
indicates that our simulations are internally self-consistent
among each others.
The figure also highlights the fact that in our simulations 
\ac{BH} collapse occurs 
for values of $\rhomax$ below the central density of the 
maximum-mass TOV star, in particular at values 
$\rho_\mathrm{max} \gtrsim 70\% \rho_\mathrm{max}^\mathrm{TOV}$
\citep{Perego:2021mkd}.
This result points to the fact that gravitational collapse is mainly
determined by the remnant core, which is slowly rotating and cold.

\subsection{Thermodynamic evolution of the remnant}
\label{ssec:thermo_remnant_evo}

\begin{figure*}
        \centering
        \includegraphics[width=0.98\textwidth]{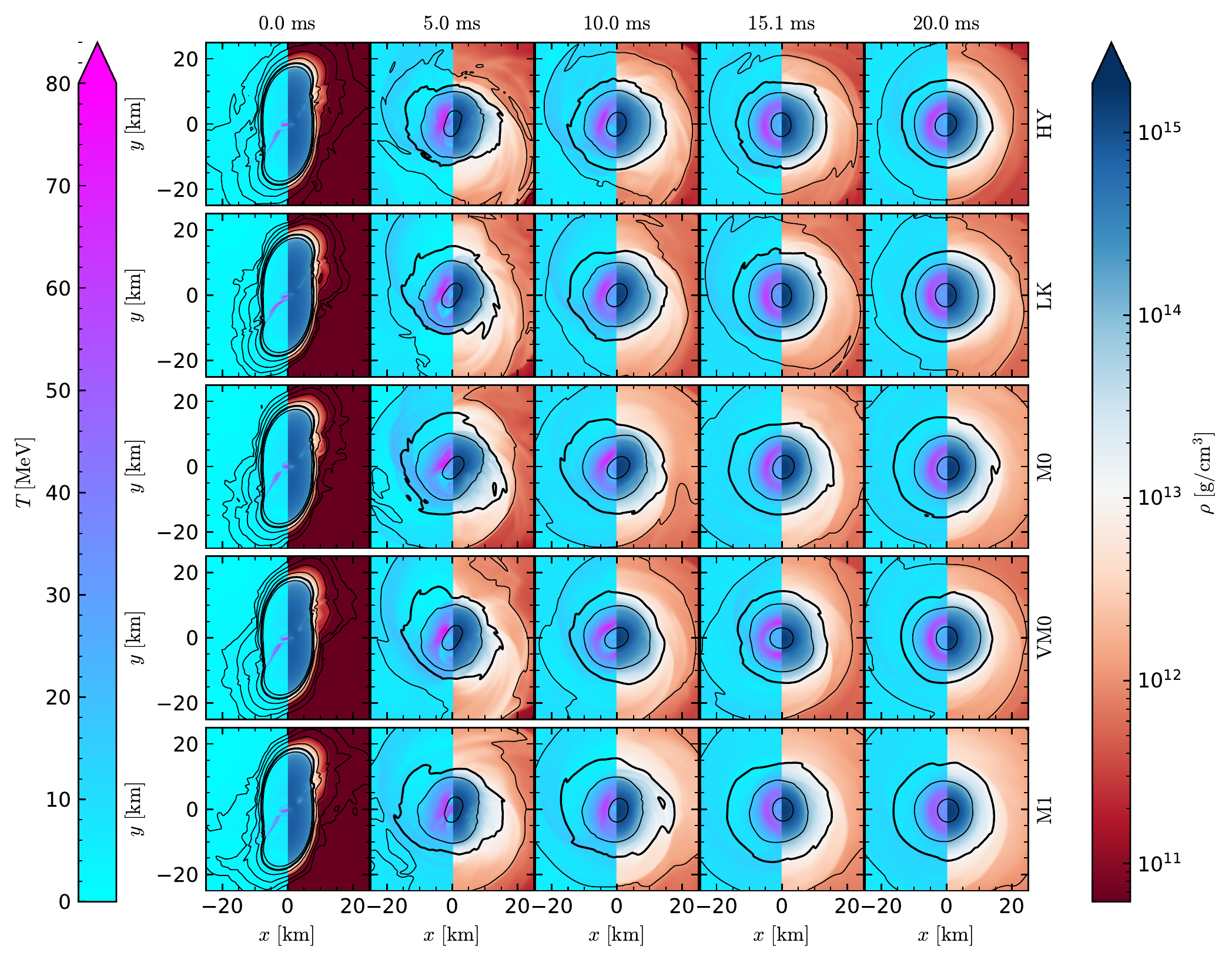}
	\caption{Comparison of the first 20 ms \pm{} evolution of
	    the remnant \ac{NS} for all LR runs.
        Each row represents a different simulation, 
        while each column corresponds to snapshots taken at the time
        expressed on top, which refers to $\ttmerg$. 
        The left half of each subplot shows the temperature
        profile, while the right half shows the density profile in
        logarithmic scale, both on the equatorial plane. 
        The black contour levels represent iso-density curves. 
        Moving away from the centre, they correspond to decreasing
        densities of $10^{15},\,10^{14},\,10^{13},\,10^{12}\dots~\rhocgs$.
        The thickest black line has density $10^{13}~\rhocgs$ and 
        conventionally denotes the interface between the remnant 
        \ac{NS} and the disc.
        }
        \label{fig:2D_LR_xy}
\end{figure*}

\begin{figure*}
        \centering
        \includegraphics[width=0.98\textwidth]{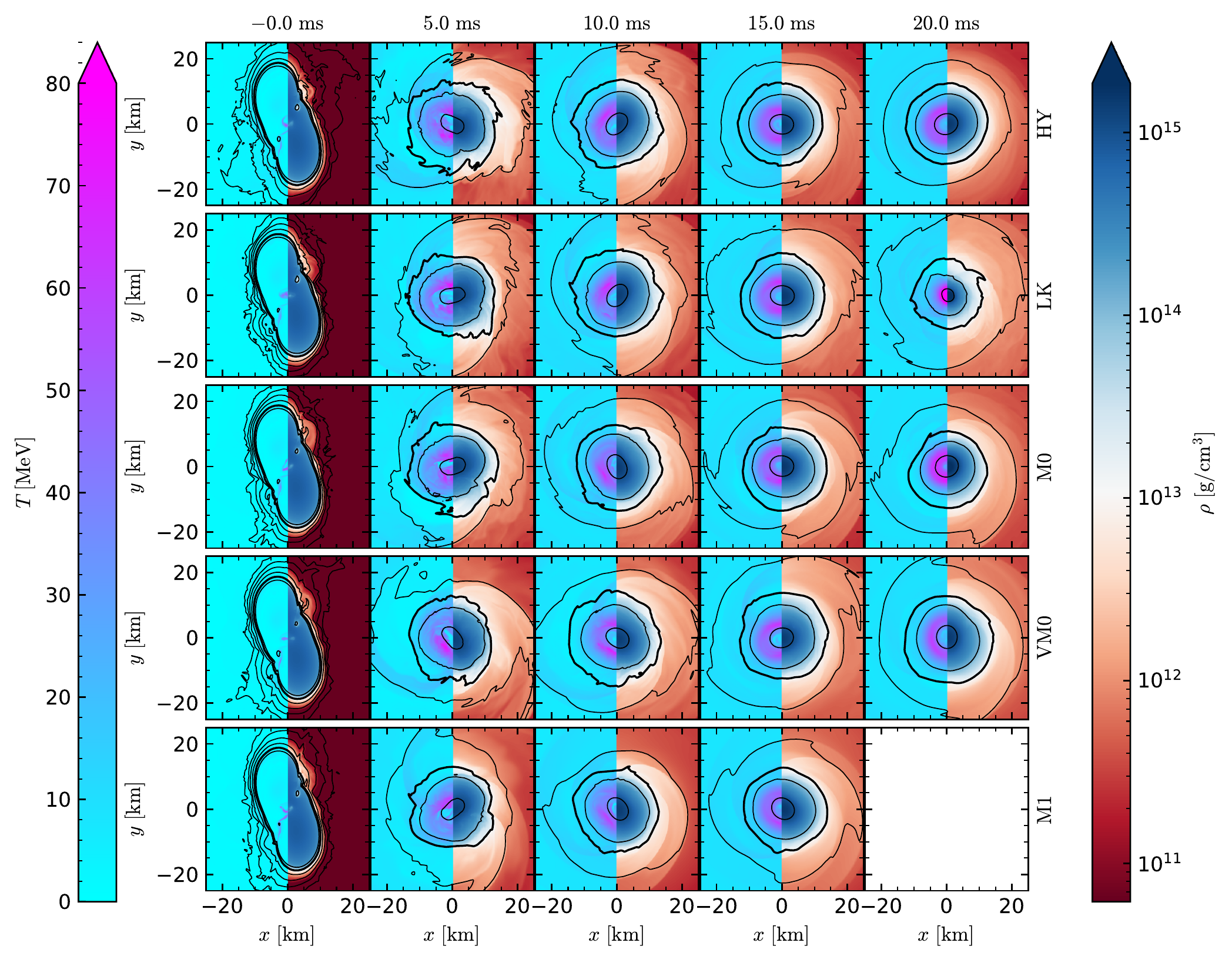}
   \caption{Comparison of the first 20 ms \pm{} evolution of
	    the remnant \ac{NS} for all HR runs.
        Each row represents a different simulation, 
        while each column corresponds to snapshots taken at the time
        expressed on top, which refers to $\ttmerg$. 
        The left half of each subplot shows the temperature
        profile, while the right half shows the density profile in
        logarithmic scale, both on the equatorial plane. 
        The black contour levels represent iso-density curves. 
        Moving away from the centre, they correspond to decreasing
        densities of $10^{15},\,10^{14},\,10^{13},\,10^{12}\dots~\rhocgs$.
        The thickest black line has density $10^{13}~\rhocgs$ and 
        conventionally denotes the interface between the remnant 
        \ac{NS} and the disc.
	    The frame corresponding to M1-HR at $\ttmerg = 20$ ms is blank 
	    because the simulation has a smaller $t_\mathrm{end}$.}
        \label{fig:2D_HR_xy}
\end{figure*}

We now discuss the impact of different neutrino schemes and
viscosity on the thermodynamics of the remnant \ac{NS}.

In Figs.~\ref{fig:2D_LR_xy} and \ref{fig:2D_HR_xy},
we report the rest-mass density and temperature profiles 
on the equatorial plane for LR and HR runs, respectively.
For both resolutions, we select snapshots at 
$\ttmerg = 0, 5, 10, 15, 20$ ms.
The remnant \ac{NS} is conventionally considered as the 
region enclosed by the iso-density shell $\rho = 10^{13}~\rhocgs$,
indicated with thick black curves in our plots.
Comparing the profiles at LR and HR, the major difference
due to resolution is that remnants at HR are more compact; this is in
agreement with the  binding energy analysis of the system in \S\ref{ssec:remnant_evo}.
The snapshots $\ttmerg = 0$ ms (first column) show the
moment in which the two \ac{NS}s touch 
and the cores start to fuse, causing the matter at the 
collisional interface to warm up because part of the kinetic energy 
is converted into thermal energy.
At 5 and 10 ms \pm{} (second and third column respectively)
the hot matter produced at the collisional interface 
forms two hotspots at peak temperature 
$T \approx 70{-}80$ MeV that revolve around the 
colder core \citep{Kastaun:2016yaf,Hanauske:2016gia,Perego:2019adq}. 
At later time $\ttmerg = 15$ ms, the hot matter
is concentrated in an annulus with a more uniform 
temperature $T \approx 60-70$ MeV.

The structure of the remnant \ac{NS} after the GW-phase is almost axisymmetric. 
The density profile decreases monotonically with the radial
coordinates, while the temperature profile does not.
In particular, the central densest region of 
$\rho \gtrsim 10^{15}~\rhocgs$ is characterised by
$T \lesssim 20$ MeV. In the region of densities 
$\rho \in [10^{14},\,10^{15}]~\rhocgs$ the temperature first
increases up to $T \approx 60-70$ MeV and then it decreases 
down to $T \approx 20$ MeV.
The layer of density $\rho \in [10^{13},\,10^{14}]~\rhocgs$
is colder, with temperatures $T \lesssim 20$ MeV. 

  With the exception of the M1 runs, which we discuss below,
we do not see any signficant differences in the remnant
density profiles comparing runs with 
different microphysics at the same resolution, as expected.
The inclusion of neutrinos emission with \ac{LK} scheme  
does not impact significantly the thermodynamics 
of the remnant's core, where matter is at high density.
Adding neutrino reabsorption with M0 scheme also does not
affect the remnant appreciably, because the component of trapped 
neutrinos is neglected and because free-streaming neutrinos mostly 
interact with the lower-density material around the remnant \ac{NS}. 
The inclusion of turbulent viscosity is also 
not expected to have a strong impact on the thermodynamics 
of the remnant core, because the effects of the viscosity
model implemented here are by construction small at densities 
higher than $10^{13}~\rhocgs$ \citep{Radice:2020ids}. In 
particular, we do not see here 
an increase in the core temperature due to kinetic energy being 
converted into thermal energy enhanced by viscosity.

\begin{figure}
        \centering
        \includegraphics[width=0.48\textwidth]{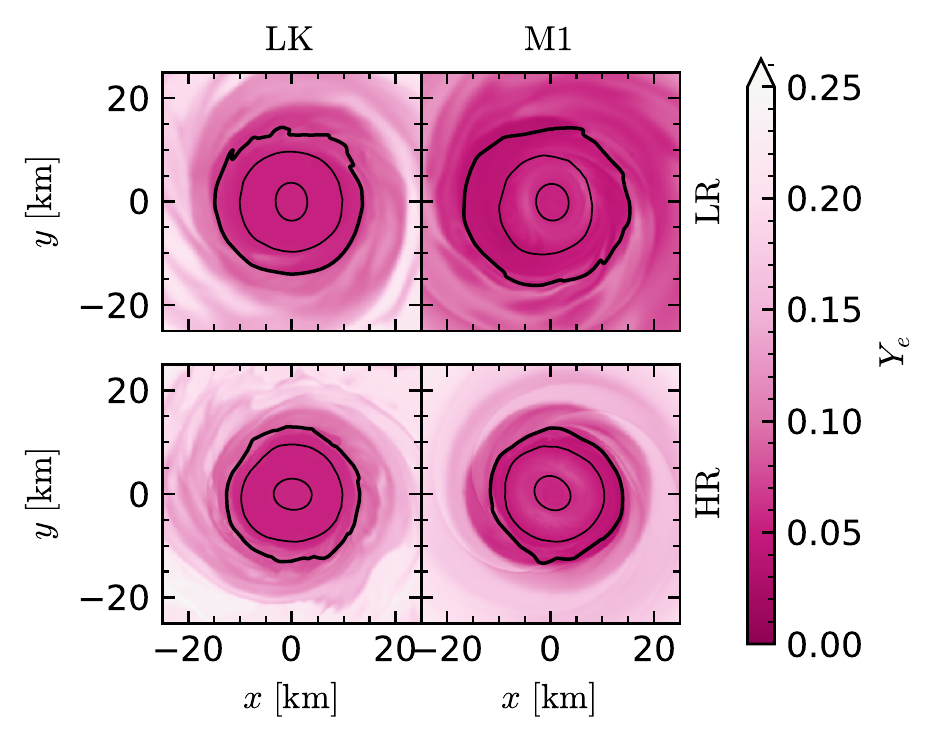}
	\caption{Comparison of $Y_e$ inside the remnant between \ac{LK} and M1 in
	         a 2D snapshot at $\ttmerg = 15$ ms on the equatorial plane.
	         The black contour levels represent iso-density curves. 
             Moving away from the centre, they correspond to decreasing
             densities of 
             $10^{15},\,10^{14},\,10^{13}~\rhocgs$.
             The thickest black line has density $10^{13}~\rhocgs$ and 
             conventionally denotes the interface between the remnant 
             \ac{NS} and the disc.
	         The remnant of the runs with M1 shows an annulus of higher
	         $Y_e$ with respect to the \ac{LK} runs at densities
	         $\rho \in [10^{14},\,10^{15}]~\rhocgs$, corresponding to 
	         the hot annuli of matter in Figs. \ref{fig:2D_LR_xy}
	         and \ref{fig:2D_HR_xy}. 
           }
        \label{fig:2D_compare_Ye}
\end{figure}

A comparison of the internal temperature of the remnant star between
the runs with \ac{LK} and M1 at 15 ms \pm{} reveals an effect due to
neutrino radiation in optically thick conditions. 
The hot annulus at densities 
$\rho \in [10^{14},\,10^{15}]~\rhocgs$ shows lower temperatures  
in the M1 run compared to the \ac{LK} case, with 
$T^\mathrm{peak}_\mathrm{M1} \approx 88 \%~T^\mathrm{peak}_\mathrm{LK}$.
This temperature difference is a physical effect due to the emergence of
a neutrino trapped gas that converts fluid thermal energy into
radiation energy \citep{Perego:2019adq}.

In Fig.~\ref{fig:2D_compare_Ye} we see the effect in the
matter composition of the remnant's core. In particular we
focus on the region $\rho \in [10^{14},\,10^{15}]~\rhocgs$ 
corresponding to the hot annulus of matter.
While in \ac{LK} runs the remnant core retains its pristine
$Y_e$ with peaks of $Y_e \approx 0.058-0.059$,
in M1 runs we report that locally $Y_e$ can be $40\%$ larger
than these values. These variations are consistent for both LR and 
HR resolutions and with Fig. 9 of \cite{Perego:2019adq}. 
The analysis of \cite{Perego:2019adq} was performed in
postprocessing from simulations without the neutrino trapped component,
finding that the presence of a neutrino gas would cause a
${\sim}33 \%$ increase in $Y_e$.
Here we confirm this effect in simulations that do simulate
the neutrino trapped component inside the remnant 
\citep[see also][]{Radice:2021jtw}.

\begin{figure}
        \centering
        \includegraphics[width=0.48\textwidth]{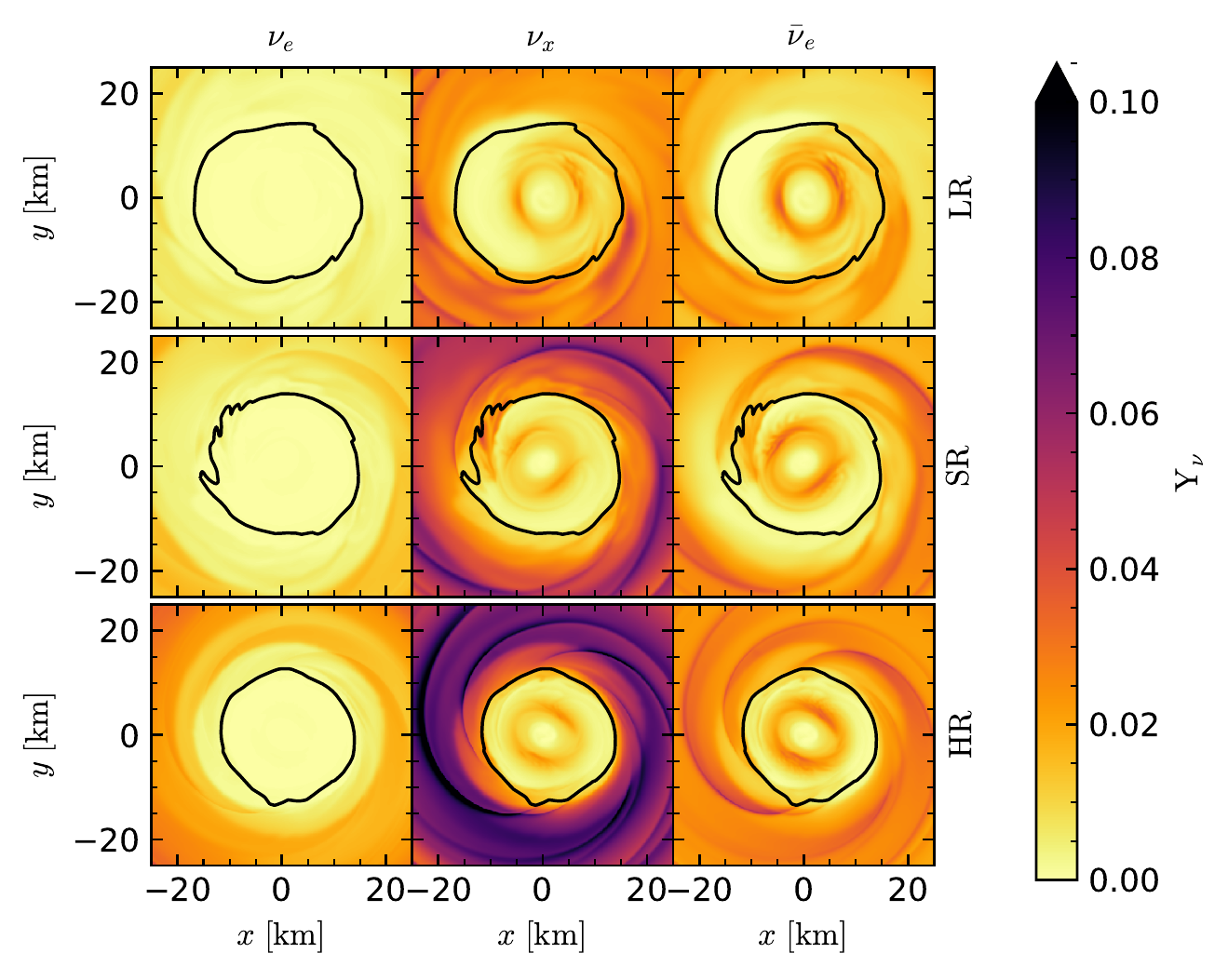}
	\caption{Neutrino fraction inside the remnant for the M1 
	simulations at $\ttmerg = 15$ ms. Each column corresponds to 
	the fraction of one of the three species simulated, 
	while each row corresponds to a different resolution.
	Inside the remnant the neutrino production 
	is favored for the species $\bar{\nu}_e$ and disfavored for 
	$\nu_e$. This finding is robust against the resolution employed.
    } 
        \label{fig:2D_compare_Ynu}
\end{figure}

Figure~\ref{fig:2D_compare_Ynu} shows 
that the thermodynamical conditions
inside the remnant is such that locally, in the high-temperature
region, the neutrino fractions follow the hierarchy 
$Y_{\nu_e} < Y_{\nu_x} < Y_{\bar{\nu}_e}$ 
\citep{Foucart:2015gaa,Perego:2019adq,Radice:2021jtw}. 
This is confirmed for all resolutions and it is explained as follows.
The matter constituting the hot annulus is characterised by 
densities $\rho \gtrsim 10^{14}~\rhocgs$ and temperatures of 
few tens of MeV. 
This is matter initially in cold, neutrino-less weak equilibrium 
coming from the collisional interface of the fusing \ac{NS} cores 
that both decompresses and heats up.
Electrons in these conditions are highly degenerate and relativistic, 
and their chemical potential ($\mu_e$) is weakly sensitive to density 
and temperature variations. On the other hand, 
neutrons and even more protons are non-degenerate, since 
their Fermi temperature 
$T_\mathrm{F}$ is such that $T \gtrsim T_\mathrm{F}$ 
and $ Y_p \sim 0.1 Y_n$ due to the initial neutron richness. The 
chemical potentials of protons ($\mu_p$) and of neutrons ($\mu_n$) are 
negative, but the absolute value of the former increases faster than the 
one of the latter. Then, the chemical potential of neutrinos at 
equilibrium, $\mu_{\nu_{e,{\rm eq}}} = \mu_p - \mu_n + \mu_e$, becomes 
negative and in particular $-\mu_{\nu_{e,{\rm eq}}} \approx 120$ MeV. 
For thermalised neutrinos in weak equilibrium, 
$\mu_{\bar{\nu}_e} = -\mu_{\nu_e}$ and 
$Y_{\nu} \propto T F_2(\mu_\nu/T)$, where $F_2(x)$ is 
the Fermi function of order 2, so that $Y_{\nu_e} < Y_{\bar{\nu}_e}$. 
Electron antineutrinos form a mildly degenerate Fermi gas, because the 
temperature is high and the degeneracy parameter $\eta_{\bar{\nu}_e} = 
\mu_{\bar{\nu}_e}/T \approx 2.5-2.7$.
Therefore, while electron neutrinos production is suppressed
due to the higher neutron degeneracy, electron antineutrinos 
production is not and a gas of $\bar{\nu}_e$ forms, with 
$Y_{\bar{\nu}_e}$ reaching peaks of ${\sim} 0.04$. 
In comparison, the maximum of $Y_{\nu_e}$ is of the order of $10^{-3}$,
while we find $\max{(Y_{\nu_x})} \approx 0.035-0.039$ 
depending on the resolution. 
This means that locally each neutrino species
constituting the effective species $x$ can be, on average, 
a factor 4 less abundant than electron antineutrinos.

\subsection{Disc evolution}\label{ssec:Disc}

After merger, part of the matter expelled during the 
collision forms an accretion disc around the remnant object.
The baryonic mass of the disc $\Discmass$ is computed from the
simulations as the volume integral of the conserved rest-mass
density
\begin{equation}
\Discmass = \int_{V} W \rho \sqrt{\gamma} d^3x,
\end{equation}
where $W$ and $\gamma$ are the Lorentz factor between
a fluid element and the Eulerian observer, and the
determinant of the spatial three metric, respectively.
In our analysis we define the disc as the baryon matter with 
density lower than $10^{13}~\rhocgs$, as in \cite{Shibata:2017xdx}.
Therefore, the integration domain $V$ extends to all the 
computational domain excluding the points inside the \ac{NS}, i.e. the region
$\rho < 10^{13}~\rhocgs$ if a massive \ac{NS} is present. 
If a \ac{BH} forms, the domain is instead restricted by excluding the 
points inside the apparent horizon using the minimum 
lapse criterion, i.e. retaining only points for which 
$\min\alpha \geq 0.3$ 
\citep[see the discussion in appendix of][for this choice]{Bernuzzi:2020txg}. $\Discmass$ for all runs are listed
in the fourth column of Tab. \ref{tab:disk_ejecta}.

\begin{figure*}
	\centering 
        \includegraphics[width=0.98\textwidth]{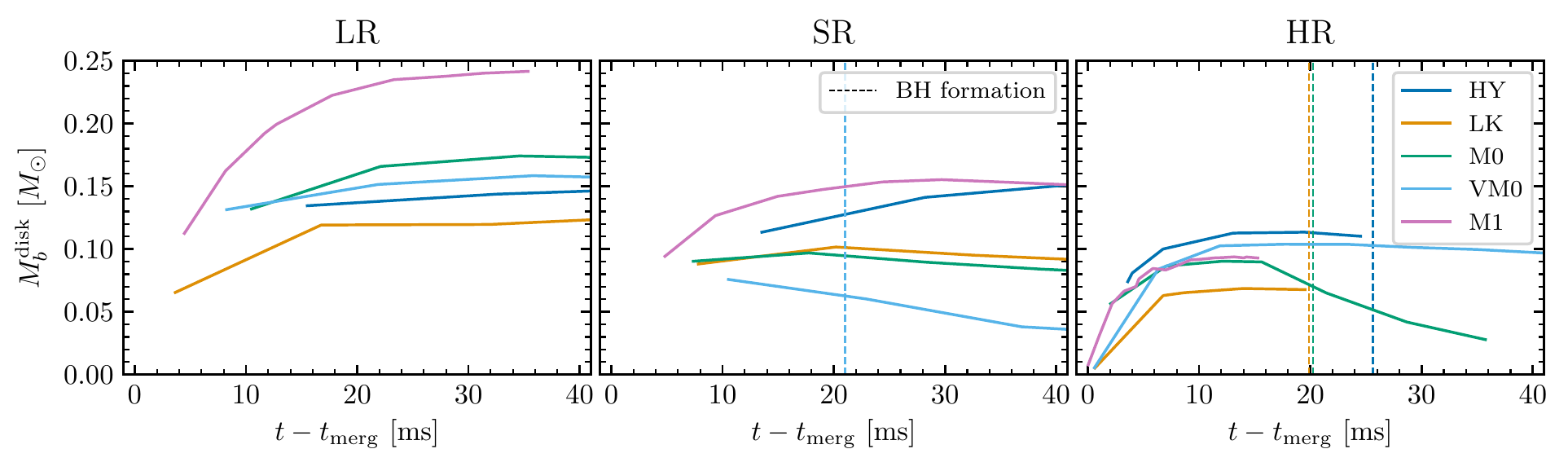}
	\caption{Time evolution of the disc mass comparing
	all runs at low, standard and high resolution. 
	The vertical dashed lines mark the time of \ac{BH} collapse.
	Time is shifted by the time of merger.}
	\label{fig:Disc_evo}
\end{figure*}

\begin{figure*}
	\centering
	\includegraphics[width=0.98\textwidth]{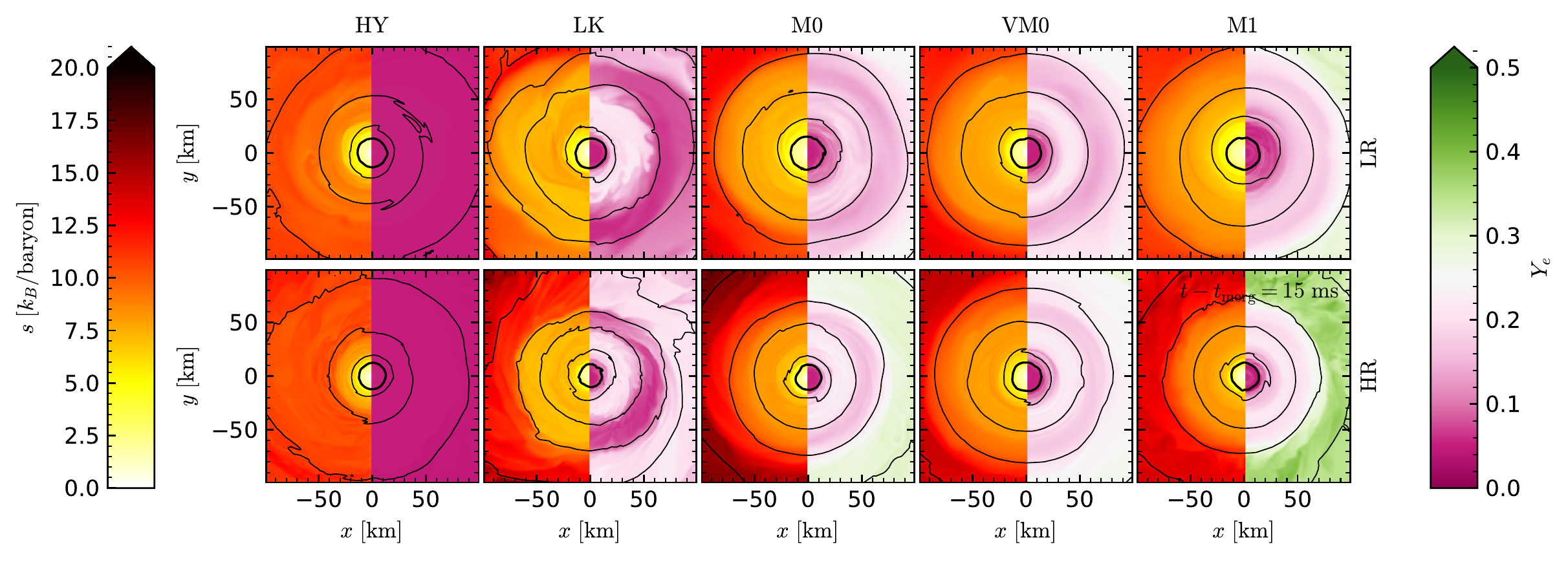}
	
	\includegraphics[width=0.98\textwidth]{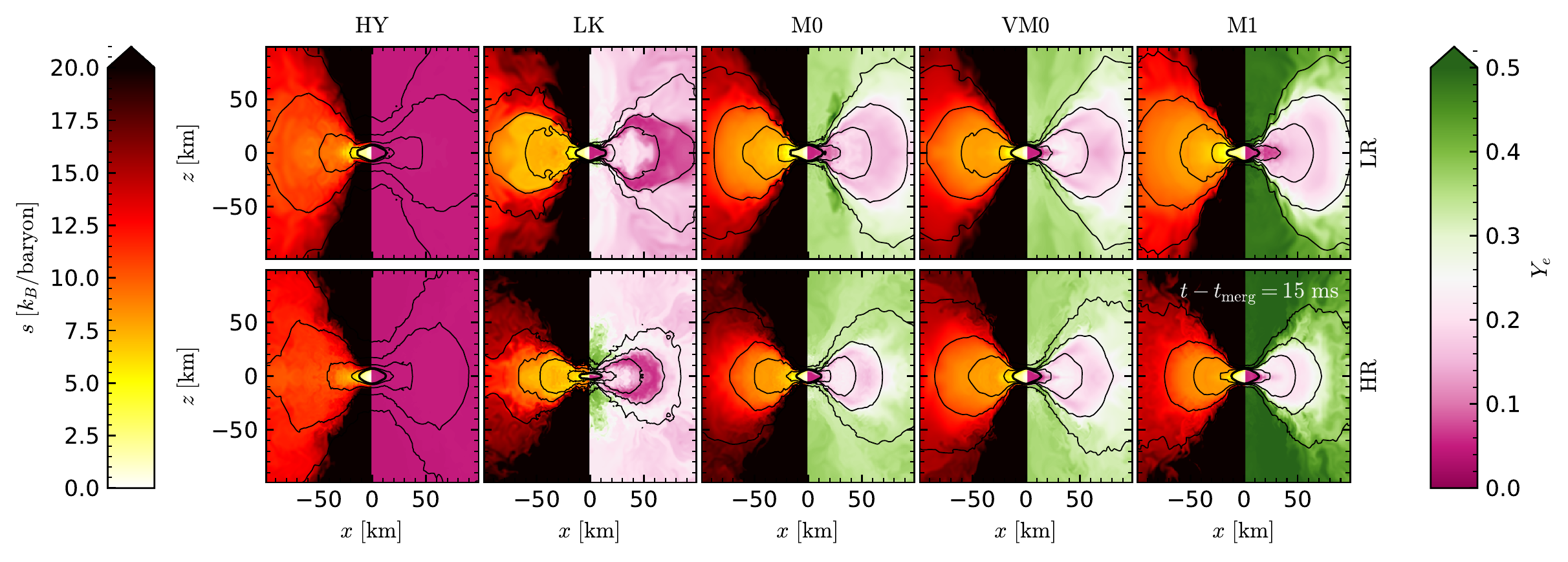}	
	\caption{2D snapshots of the $xy$-plane (top plot) and
	         of the $xz$-plane (bottom plot) showing the properties
	         of the accretion disc around the \ac{NS} remnant 
	         at $\ttmerg = 20$ ms. In each plot we compare all runs 
	         at LR (top row) to all runs at HR (bottom row). 
	         Each frame shows the matter specific entropy $s$ 
	         on the left half and the electron fraction
	         on the right half. The thickest black curve is the isodensity
	         contour $\rho = 10^{13}~\rhocgs$ delimiting the \ac{NS}
	         remnant, while the other thinner curves 
	         moving outwards correspond to densities 
	         $\rho = 10^{12},\,10^{11},\,10^{10},\dots~\rhocgs$.
	         Note that the profiles for M1-HR run are reported at 
	         15 ms \pm{}, close to the last available time.
	         	         	         }
	\label{fig:Disc_2d}
\end{figure*} 

In Fig. \ref{fig:Disc_evo} we report the first 40 ms \pm{} 
time evolution of $\Discmass$. The largest increase in the disc mass
happens within ${\sim} 10$ ms \pm{}, as a result
of the collision and of the successive bounces of the two merging
cores. On this timescale the mass of the disc reaches values of the
order of ${\sim} 0.1~\Msun$, and then it 
stays constant for a few tens
of ms, if the remnant \ac{NS} does not collapse. 
When a \ac{BH} forms, the disc mass drastically drops because
a large fraction of the disc is swallowed by the \ac{BH}.

To discuss the differences due to microphysics we
focus on HR runs. In HY-HR, when neutrinos are not simulated, 
the disc mass is the largest, being almost double the LK-HR one.
Even before \ac{BH} formation, \ac{LK} run exhibits the smallest disc mass
among all the runs, with $\Discmass \approx 0.06\,\Msun$.
This can be explained by the fact that \ac{LK} cools down 
the lower-density matter around the \ac{NS} core, causing 
the outer shells of the remnant \ac{NS} to be 
less inflated and to expell less matter.
When neutrino reabsorption is present (M0-HR, M1-HR) the disc mass increases
to $\Discmass \approx 0.09\,\Msun$ and is very similar among the two runs.
For VM0-HR, angular momentum and
matter transport enhanced by viscosity has the effect of
increasing the disc mass with respect to M0 only. Eventually 
$\Discmass$ reaches an intermediate value between HY-HR and M0-HR ones.

We observe a systematic dependence on resolution in the amount 
of disc mass.
LR runs present the largest $\Discmass$ for all simulations.
Here, the minimum mass is found for LK-LR run, 
with ${\sim} 0.12~\Msun$, while in M0-LR, VM0-LR and HY-LR runs
$\Discmass$ reaches similar masses ${\sim}0.15,\,\Msun$. The
largest disc mass is obtained for M1-LR simulation, with almost 
${\sim} 0.25\,\Msun$. 
For this resolution a stable rotating \ac{NS} forms and we also 
observe that
the disc mass slowly increases with time on timescales longer than the ones shown in the plot. This is due to the fact that  
some matter is explled from the outer shell of the remnant \ac{NS} 
and becomes part of the disc \citep{Radice:2018xqa}.
For increasing resolution the disc mass decreases comparing each run
with its lower resolution counterparts, except for HY-SR. The decrease
can be as large as $44\%$ (M0-LR vs. M0-SR). 
HR runs show the smallest $\Discmass$. 

Finite resolution also impact the disc mass indirectly by determining
different collapse times. 
Higher resolution simulations can predict final 
disc masses that are much smaller than lower resolution ones
when a \ac{BH} forms and it swallows part of the disc.
The presence of such lighter discs due to \ac{BH} collapse
can have a large impact on the emission of gravitationally 
unbound material from the disc at secular timescales 
\citep[see, e.g.,][]{Camilletti:2022jms,Radice:2018pdn}. 
We note however that, as long as gravitational 
collapse does not occur, the spread of $\Discmass$ 
due to different microphysics is smaller as the resolution increases.

In Fig. \ref{fig:Disc_2d} we compare the geometric properties and
the composition of the disc among the LR and HR runs as 2D snapshots
of the $xy$-plane (top plot) and $xz$-plane (bottom plot)
at $\ttmerg = 20$ ms. 
The geometry of the disc can be analysed by means of the black iso-density 
contours in the figure.
The high-density portion of the disc
$\rho \in [10^{12},\,10^{13}]\,\rhocgs$  extends to ${\sim} 20$ km
in the equatorial plane and ${\sim} 10$ km in the $xz$-plane.
The region $\rho \in [10^{11},\,10^{12}]~\rhocgs$ is more inflated when 
neutrinos are present, compared to the HY case, in both $xy$- and $xz$-
planes.
The low-density $\rho \approx 10^{10}~\rhocgs$ tails of the disc extends up 
to tens of km from the central object on the equatorial plane.

The most evident difference among resolutions is that discs 
are geometrically smaller for higher resolutions. If we consider
the iso-density curve $\rho = 10^{10}~\rhocgs$ on the 
orbital plane, it extends to ${\sim} 90$ km for LK-LR and ${\sim} 65$ km
for LK-HR. Similar numbers are found for M0 runs, while in  VM0 runs
the difference between LR and HR is smaller, ${\sim} 10$ km. The
largest difference is found in M1 runs, for which the curve extends to
$\gtrsim 100$ km for LR and to ${\sim}65$ km at HR.

For the composition of the disc we refer to the the entropy and electron 
fraction profiles in Fig. \ref{fig:Disc_2d}.
The high-density matter $\rho \in [10^{12},\,10^{13}]\,\rhocgs$ is
characterised by low electron fraction and low entropy because it is 
made of fresh matter expelled from the remnant \ac{NS}.
In the HY runs the electron fraction is frozen at $Y_e = 0.05$
because neutrinos are not simulated.
Comparing HY (left column) in the bottom plot with 
the others, we note that the presence of neutrinos clears 
the polar regions above the remnant 
NS \citep{Radice:2016dwd,Mosta:2020hlh}. 
In the runs with neutrinos, at this density,  
we see that $Y_e$ increases up to values $\gtrsim 0.2$ which indicates
that matter protonises.
In \ac{LK} runs, for decreasing density and increasing distance
from the remnant the $Y_e$ first increases as mentioned above,
then decreases to $\sim 0.1$ at $\rho \in [10^{10},\,10^{11}]~\rhocgs$.
At lower densities and high latitude $Y_e \lesssim 0.25$.
We note that in the region right above the remnant $Y_e \approx 0.4$, at LR.
At HR the remnant is close to \ac{BH} collapse and this causes a
temperature increase and consequently an increase of electron fraction 
in the low-density matter above the remnant.
The M0 and VM0 runs show different disc composition with respect to
LK but similar between each others. Here, 
a fraction of neutrinos streaming out of the \ac{NS}
remnant is absorbed by lower-density material, 
increasing its $Y_e$.
$Y_e$ in the shell $\rho \in [10^{10},\,10^{11}]~\rhocgs$ 
is larger than in the \ac{LK} case at the same density. 
For increasing latitudes (and decreasing density) $Y_e$ increases,
reaching values up to $Y_e \approx 0.35$.
In the M1 runs the electron fraction has larger values when 
comparing shells of same density
to M0 or VM0. In particular matter at high latitude and low density 
reaches $Y_e \approx 0.5$, and is thus quantitative different from M0 runs.
The features described are robust against resolution changes.

\section{Gravitational waves}
\label{sec:gws}

\begin{figure*}
	\centering 
        \includegraphics[width=0.98\textwidth]{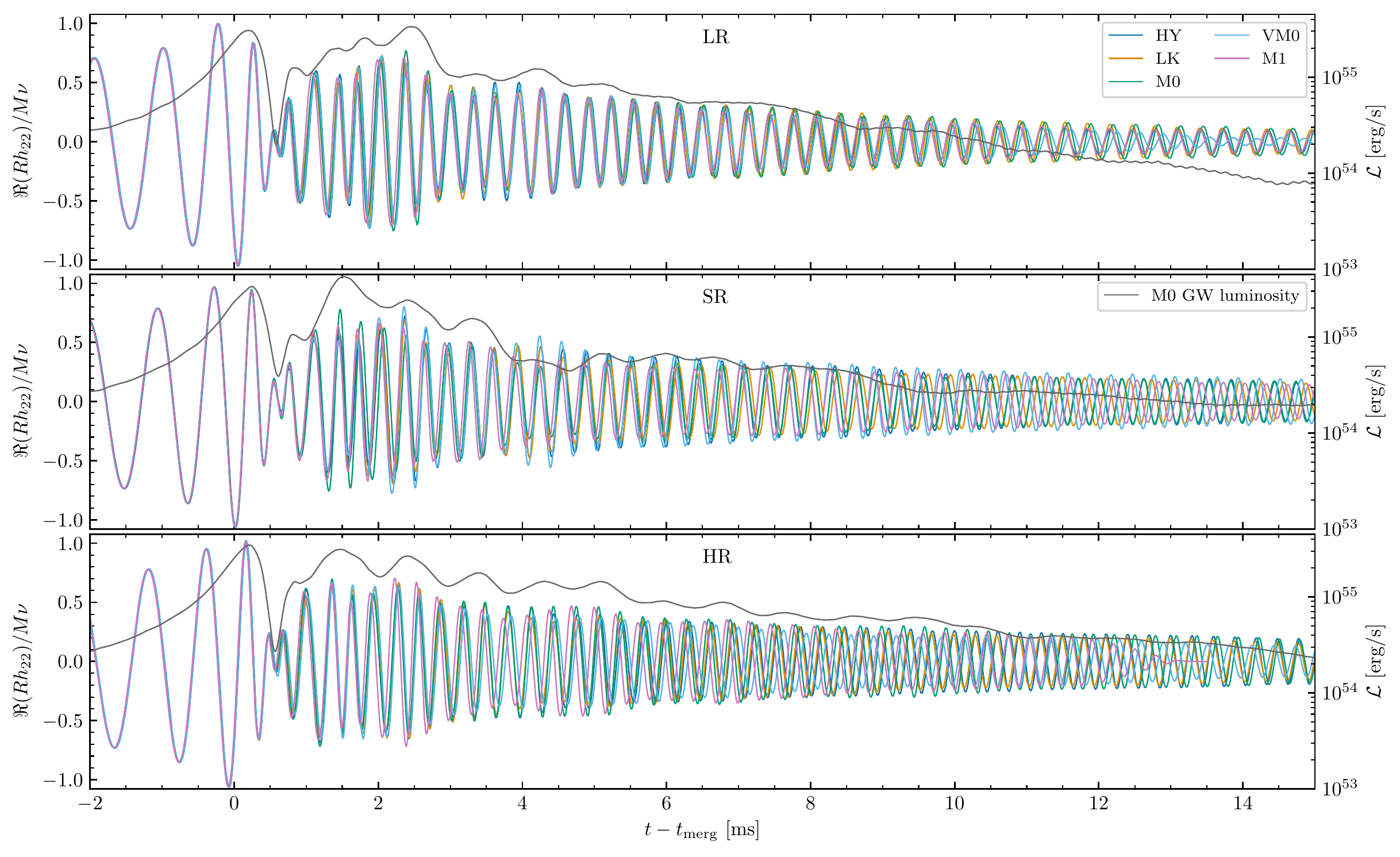}
	\caption{Real part of the $(\ell,\,m) = (2,\,2)$ mode of 
	         the GW strain normalised by the total mass of the
	         system $M$, the symmetric mass ratio 
	         $\nu := M_1M_2/(M_1 + M_2)^2$ and the                  
	         extraction radius $R$.
	         The grey curve corresponds to one representative 
	         GW luminosity	curve (we choose here the M0 runs).
	         Time is shifted by the merger time.}
	\label{fig:strain}
\end{figure*}

We now compare the \ac{GW}s emitted during the \ac{BNS} merger
in our simulations.
The modes of the gravitational wave strain $h_{\lm}$ are computed
from the Weyl scalar $\Psi_4$ projected on coordinate spheres and decomposed in $s=-2$ spin weighted spherical harmonics, $\psi_{\lm}$.
We solve
\begin{eqnarray}
\label{eq:strain}
\psi_{\ell m}=\ddot{h}_{\ell m}\,,\label{eq:hlm}
\end{eqnarray}
using the method of \cite{Reisswig:2010di}; the strain is then given by the mode-sum:
\begin{equation}
  R\left(h_+ - i h_\times\right) = \sum_{\ell=2}^{\infty}\sum_{m=-\l}^\l h_{\ell m}(t)\; {}_{-2}Y_{\ell m}(\vartheta,\varphi)\,.
\end{equation}
where $R$ is the finite extraction radius in our simulations.
Following the convention of the LIGO algorithms library \citep{lalsuite,Cutler:1994ys}, we let
\begin{equation}\label{eq:strain_complex}
R h{}_{\ell m} = A{}_{\ell m} \exp(-i\phi{}_{\ell m}),
\end{equation}
and compute the gravitational-wave frequency as
$\omega{}_{\ell m} = d\phi{}_{\ell m}/dt$, $f_{\ell m} = \omega{}_{\ell m}/2\pi$.
In Fig.~\ref{fig:strain} we compare the
$(2,\,2)-$ mode of the GWs among our runs up to 16 ms \pm{}. 
We additionally report the GW luminosity 
$\L_\mathrm{GW} := dE_\mathrm{GW}/dt$
for one representative run (M0 for every resolution).
Up to merger, the waveforms do not show any significant differences among
each others. The amplitude peaks at 
$R A_{22}^\mathrm{merg}/M\nu \approx 1.06$ with a merger frequency of $f^\mathrm{merg}_{22} \approx 1.9$ kHz. The \pm{} spectrum peak frequency is $f_2 \approx 3.2$ kHz. 
These three quantities are measured quite robustly from 
our simulations.
At LR, the maximum variations of $A_{22}^\mathrm{merg}$, 
$f^\mathrm{merg}_{22}$ and $f_\mathrm{2}$
among all the runs are respectively 
${\sim} 0.3\%$, ${\sim} 1.3\%$, ${\sim} 2.1\%$.
At SR the maximum variations of these quantities are below 
$0.7 \%$.
Lastly, for HR runs the maximum variations of $A_{22}^\mathrm{merg}$, 
$f^\mathrm{merg}_{22}$ and $f_2$ among all the runs are respectively 
${\sim} 0.38\%$, ${\sim} 1.1\%$, ${\sim} 2.1\%$.
The differences due to finite resolution are instead generally larger.
We find maximum differences between SR and HR of ${\sim} 1.3 \%$ for $A_{22}^
\mathrm{merg}$, ${\sim} 4.1 \%$ for $f_{22}^\mathrm{merg}$ and 
${\sim} 2 \%$ for $f_2$.
The GW luminosity peaks shortly after merger at
$\L_\mathrm{GW}^\mathrm{peak} \approx 3.5 \times 10^{55}$~erg~s$^{-1}$,  consistently with
\cite{Zappa:2017xba}. The peak value has a maximum variation of 
${\sim} 34 \%$ among our runs.

In the \pm{} waveform we see significant differences in the amplitude,
frequency and phase evolution among the runs. 
We first analyze phase convergence among different resolutions and 
fixed microphysics prescription, and obtain approximately first order 
convergence. Then, in order to study the impact of resolution over the 
different simulated physics, we perform a faithfulness analysis between 
pairs of waveforms. The faithfulness between two waveforms $h_1(t)$ and 
$h_2(t)$ is defined as
\begin{equation}
\mathcal{F} := \max_{t_c,\,\phi_c} \frac{(h_1|h_2)}{\sqrt{(h_1|h_1)(h_2|
h_2)}}
\end{equation}
where $t_c,\,\phi_c$ are the time and phase of the waveforms 
at a reference time, the Wiener inner product is
\begin{equation}
(h_1|h_2) := 4\Re \int\frac{{\tilde{h}_1(f)\tilde{h}_2^*(f)}}{S_n(f)}df\,,
\end{equation}
the symbol $\sim$ denotes the fourier transform, and $S_n(f)$ is 
the power spectral density of the Einstein Telescope.
The unfaithfulness is defined as the complement, $\overline{\mathcal{F}} := 1- \mathcal{F}$.
In the context of GW parameter estimation, two waveforms are distinguishable if their faithfulness satisfies the necessary criterion 
\citep{Damour:2010zb}
\begin{equation}
\mathcal{F} > 1 - \frac{\epsilon^2}{2\varrho^2}\,,
\end{equation}
where $\varrho$ is the matched-filtered signal to noise ratio (SNR) and we take $\epsilon^2 = N$, with 
$N$ number of intrinsic parameters of the system \citep{Chatziioannou:2017tdw}.
From the above inequality, the minimum SNR that allows to detect the differences between two waveforms can be estimated as
\begin{equation}\label{eq:SNR}
\varrho \approx \sqrt{\frac{N}{2 \overline{\mathcal{F}}}}.
\end{equation}

We compare all our runs in pairs, in such a way that the two runs
in a pair have either the same resolution (e.g. HY-LR and LK-LR) or
the same microphysics
(e.g. HY-LR and LK-LR), excluding comparisons of the kind 
HY-LR and LK-SR. 
At LR, we find a maximum mismatch of 
$\overline{\mathcal{F}} \approx 0.087$
between HY-LR and M0-LR runs. At SR the mismatches are generally larger and we obtain 
a maximum value of $\overline{\mathcal{F}} \approx 0.2$ 
between LK-SR and M1-SR and also between M0-SR and VM0-SR runs. At HR the mismatches are the largest and we obtain a maximum of 
$\overline{\mathcal{F}} \approx 0.32$ between the VM0-HR and M1-HR runs.
Comparing runs at different resolutions, we obtain 
mismatches of the order of few times $10^{-1}$ in almost all
comparisons. The only two exceptions are HY-LR vs. HY-SR and 
LK-LR vs. LK-SR for which $\overline{\mathcal{F}}$ is 
few times $10^{-2}$.

Our analysis indicates that possible effects due to 
microphysics can be detected in the GW signal
only in the \pm{}.
However, GW models used for matched filtering that are
informed on NR simulations~\citep{Breschi:2019srl,Breschi:2022xnc} 
at LR would not be accurate enough
to detect such effects. 
In particular, differences due to the simulations' finite resolution 
would be dominant in such GW models. 
At SR and HR, mismatches between waveforms of runs performed 
with different microphysics are
comparable to the ones due to finite resolution.
GW templates constructed with these data might be able to 
distinguish such differences in the signal from $\varrho \gtrsim 4$ 
(Eq.~\eqref{eq:SNR}). Notably, this precision might be sufficient 
for third generation observations, since differences
in the signals due to variations in the \ac{EoS} 
at extreme matter densities are potentially
observable at \pm{} SNR ${\sim} 8$ \citep{Breschi:2022xnc}.

Our results indicate that simulations at SR or HR are necessary in order 
to distinguish differences due to microphysics in the remnant. 
In particular, our high-resolution M1 simulations do not show any 
evidence for significant out-of-equilibrium and bulk viscosity 
effects in the waveforms. This is in agreement with the findings of
\cite{Radice:2021jtw} that were obtained at LR, but it is in constrast 
with Refs.~\cite{Most:2022yhe,Hammond:2022uua}. 
The simulations performed for the latter works do not consider weak 
interactions or use a \ac{LK} scheme and are performed at a maximum resolution
of 400 m, which is much lower than our LR.

\section{Mass ejecta}\label{sec:ejecta}

We analyze the material ejected on dynamical timescales and up to
${\sim}20$~ms \pm{}. These ejecta include the full dynamical
ejecta component and the early portion of the spiral-wave wind
component.
The dynamical ejecta is composed of a tidal component originating from
tidally unbound NS material and a shocked component originating from
the first bounce after the core collision \citep{Radice:2018pdn}.
The tidal component is launched mostly across the equatorial
plane and is characterised by a low $Y_e\approx 0.05-0.15$
and low entropy,
$s \lesssim 5~k_\mathrm{B}~\mathrm{baryon}^{-1}$.
The shocked component has higher entropy than the tidal component and 
peak temperature of tens of MeV, which produces large amount of
electron-positron pairs with consequent increase of $Y_e$ due to
positron captures on neutrons. 
Neutrino irradiation from the remnant can further increase
$Y_e$ of this ejecta component
through absorption on neutrons, especially at high latitudes where 
neutrino emission is more efficient. The shock-heated ejecta 
expand over the entire solid angle due to interaction with 
the tidal ejecta, hydrodynamics shocks and weak interaction,
with a preference for the emission on the equatorial plane.

Other mechanisms can unbind material from the disc and they act 
generally on longer timescales. Spiral-wave winds can originate from 
non-axisymmetric density waves from the NS remnant
~\citep{Nedora:2020pak}. The remnant's spiral 
arms transport angular momentum outwards in the disc and material gets
then unbound from the disc edge.
On longer timescales disc winds can develop, also powered by neutrino
reabsorption~\citep[e.g.][]{Dessart:2008zd,Perego:2014fma,Just:2014fka,Fujibayashi:2017xsz,Rosswog:2022tus} but our
simulations are not sufficiently long to capture this component.

\begin{figure*}
	\centering
	\includegraphics[width=0.98\textwidth]{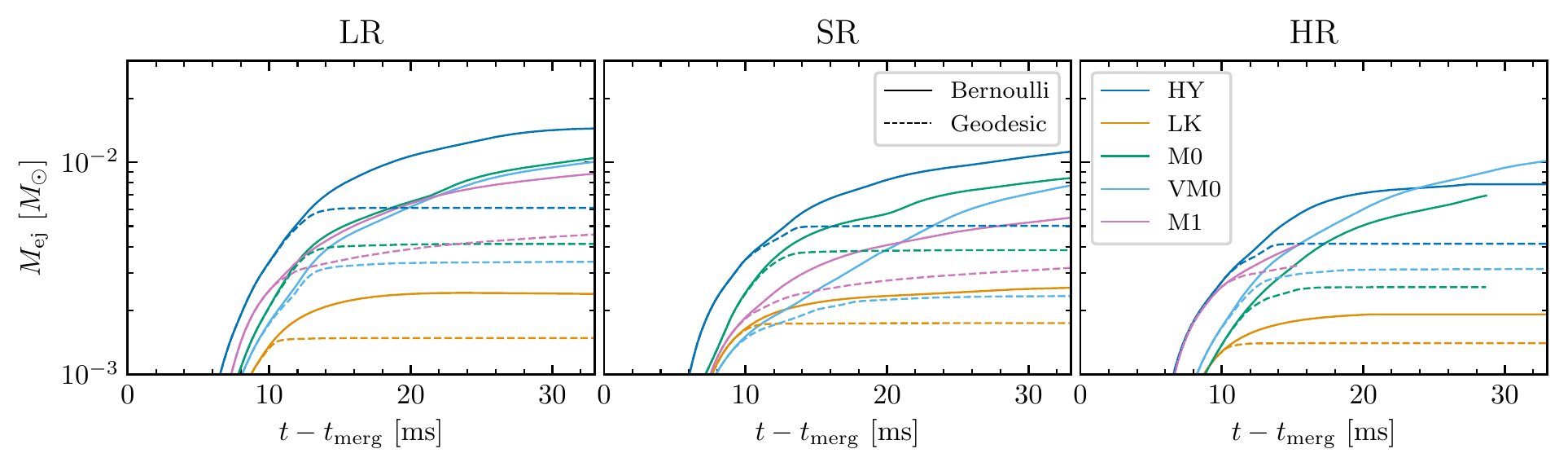}
	\caption{Time evolution of the ejecta mass extracted at $R=443$ km
	comparing the Bernoulli and geodesic criteria. 
	Mass is reported in logarithmic scale and compared across
	resolution. Within each subplot we compare the ejecta 
	for all runs.}
	\label{fig:ejecta_mass}
\end{figure*}

In the literature there are two main ways to identify the unbound material
from simulations, namely the geodesic and the Bernoulli criterion 
\citep[see, e.g.,][for a recent work on this topic]{Foucart:2021ikp}.
The geodesic criterion assumes that ejecta follow spacetime geodesics
in a time-independent, asymptotically flat spacetime. 
Therefore, a particle is considered unbound if $u_t < -1$,
where $u_t$ is the time component of the particle's 4-velocity.
According to the Bernoulli criterion, a
fluid element is considered unbound if $h u_t \leq -1$, 
where $h$ is the fluid specific enthalpy, 
$h = 1 + \epsilon + p/\rho$. 
Here $\epsilon$ is the specific internal
energy, and $p$ and $\rho$ are the pressure and rest-mass 
density of the fluid, respectively. 
The asymptotic velocity of the unbound particle is calculated as 
$v_\infty \simeq \sqrt{2\left(h \left( E_\infty +1\right)-1\right)}$. 
This criterion assumes that $hu_t$ is constant along a streamline of
a steady-state flow. This assumption is correct if the metric and 
the flow are both stationary.
Even though this is not formally true for
merger outflows, this criterion is considered sufficient to account 
for the gain in kinetic energy of the expanding matter in 
the outflow due to thermal and nuclear binding energy 
\citep{Foucart:2021ikp}.
The geodesic and Bernoulli criteria can be used to conventionally
identify (separate) the dynamical ejecta from the wind ejecta \citep{Nedora:2021eoj}.

In Fig.~\ref{fig:ejecta_mass} we present the evolution of the ejecta 
mass in our simulations, comparing the geodesic and Bernoulli criteria.
At 20 ms \pm{} the ejecta masses calculated with the geodesic criterion 
are saturated, except for the M1 runs. As expected, the ejecta 
mass calculated with the Bernoulli criterion is larger than the
one estimated with the geodesic criterion at 
comparable times. The ejecta mass in the Bernoulli case
keeps increasing at later time due to the 
contributions of the spiral-wave winds. 
In the rest of this section we refer to and discuss the Bernoulli ejecta.

The ejecta mass shows a steep increase up to ${\sim}10$ ms 
\pm{} in all the runs and then it tends to saturate at few
tens of ms after merger. Within ${\sim}20$ ms \pm{} a mass
of $\gtrsim 2 \times 10^{-3}~\Msun$ is typically ejected. 
We refer to Tab. \ref{tab:disk_ejecta} for the quantitative values
at a fixed time for all the runs.
For HY runs, $\gtrsim 8 \times 10^{-3}~\Msun$ of matter is expelled,
which represents the largest matter emission among all the runs.
The ejecta mass in \ac{LK} runs is systematically one order of 
magnitude lower than that of all the other runs, consistently 
with \cite{Radice:2016dwd}. 
This happens because the neutrino cooling reduces the enthalpy of
the material and as a result the emission is largely decreased.
When neutrino reabsorption is included through the M0 scheme, the effect
of cooling is counteracted by the neutrino energy deposition in the
shock-heated ejecta and $\mej$ becomes larger 
than the \ac{LK} case, reaching values $\gtrsim 10^{-2}~\Msun$.
The evolution of $\mej$ in VM0 runs follows a similar behaviour.
$\mej$ measured in M1 and M0 runs are comparable, within 
a few tens of percent.

Focusing on the effects of finite resolution, 
we observe a monotonic decrease
of $\mej$ for increasing resolution for all the runs.
Since the onset of \ac{BH} collapse stops the matter ejection,
we measure smaller final ejecta masses in HR simulations
than the other cases. Comparing the variations in the ejected
mass at a fixed time of 20 ms \pm{} due to resolution,
we obtain a maximum variation of ${\sim}50 \%$ between 
VM0-SR and VM0-HR.

\begin{figure*}
	\centering
	\includegraphics[width=0.98\textwidth]{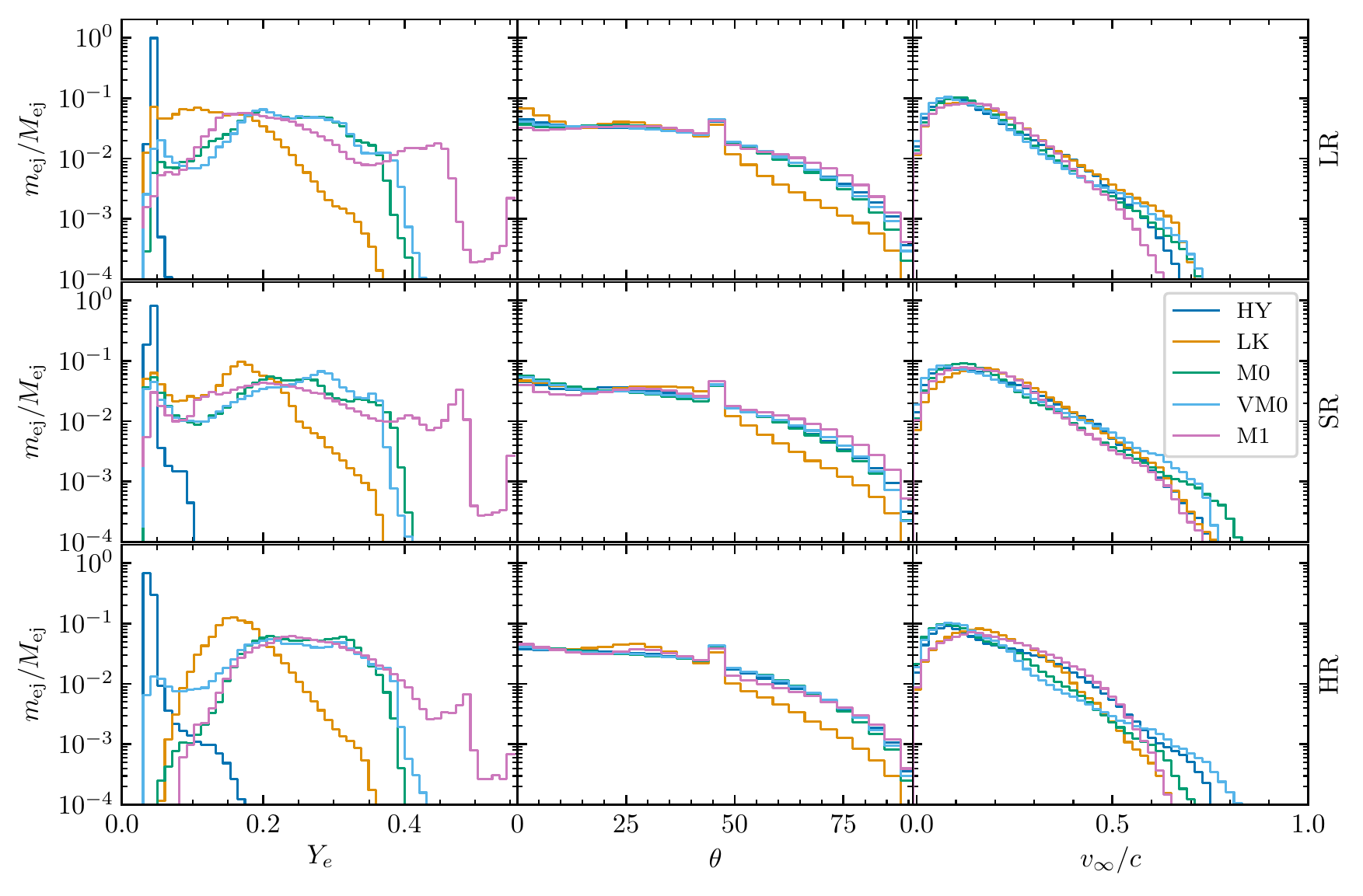}
	\caption{Histograms of the ejecta extracted at $R=443$ km. 
	Each row shows the fraction of ejecta mass
	in a bin normalised to the total ejecta mass
	in logarithmic scale for different resolutions. 
	In each columns are represented, respectively, 
	the electron fraction, 
	the latitudinal distribution
	and the asymptotic velocity of the ejecta. In each frame
	we compare the ejecta properties among all runs.
	We perform the analysis
	until $\ttmerg \approx 20$ ms, corresponding to the 
	earliest $t_\mathrm{end}$ of our set of simulations 
	(see Tab. \ref{tab:disk_ejecta}). For M1-HR we use 
	the last available time $t_\mathrm{end} \approx 15$ ms \pm{}.}
	\label{fig:ejecta_hist}
\end{figure*}
The most salient properties of the ejected material are summarised in the 
histograms of Fig.~\ref{fig:ejecta_hist}. We stress that the histograms 
produced using the geodesic criterion do not significantly differ from those  
obtained with the Bernoulli criterion that we show here.
In Tab. \ref{tab:disk_ejecta} we report mass-weighted averages of the 
same quantities presented in the figure.
Most of the mass is emitted almost uniformly
in the interval $0^\circ \leq \theta \lesssim 50^\circ$ (second column
of Fig.~\ref{fig:ejecta_hist}). 
The peak at $\theta \approx 45^\circ$ is due to an artefact in the 
mass extraction and it is not physical.
At larger angles, the mass emission is slightly more suppressed in 
\ac{LK} runs with respect to the other cases \citep{Radice:2016dwd}.
The average emission angle for all runs is enclosed in
$\theta \in [27^\circ,\,37^\circ]$ and
is systematically lower for \ac{LK} at all resolutions.

The asymptotic velocity distribution is peaked around values in the interval
$0.15 \leq v_\infty/c \lesssim 0.22$. The velocity distribution has fast 
tails reaching ${\sim}0.8$~c. These tails can originate a 
radio-X-ray afterglow to the kilonova emission, peaking at years \pm{} 
timescales 
\citep{Nakar:2018uwt,Hotokezaka:2018gmo,Hajela:2021faz,Nedora:2021eoj}. 
We measure a mass in the fast 
tail of the ejecta, i.e. with asymptotic velocity 
$v_\infty/c \geq 0.6$, of ${\sim} 10^{-6}-10^{-5}~\Msun$
(see Tab.~\ref{tab:disk_ejecta}).

We find that it is possible to model the function 
$M_{\rm ej}(v_\infty/c)$ approximately with 
a broken power law of the kind
\be
  M = M_0 \begin{cases}
    \left(\frac{\beta\gamma}{(\beta\gamma)_{\beta_0}}\right)^{-s_\mathrm{KN}} & 0.1 < \beta\gamma < (\beta\gamma)_{\beta_0}\\
    \left(\frac{\beta\gamma}{(\beta\gamma)_{\beta_0}}\right)^{-s_\mathrm{ft}} & \beta\gamma > (\beta\gamma)_{\beta_0}
  \end{cases}
\ee
where $\beta = v/c$, $\gamma$ is the corresponding Lorenz factor and
$(\beta\gamma)_{\beta_0}=\beta_0\cdot \gamma(\beta=\beta_0)$.
The values of $\beta_0$ defining the ``breaks'' in the broken power
vary in the range ${\sim}0.3-0.45$.
Fitting parameters are $M_0 \approx (3.2-17) \times 10^{-5}~\Msun$, 
$s_\mathrm{KN} \approx 0.64-1.6$ and the ejecta tail with 
$v_\infty/c \gtrsim \beta_0$ can have a rather steep dependence 
on the velocity, with $s_\mathrm{ft} \approx 4-11$.

\begin{figure*}
	\centering
	\includegraphics[width=0.98\textwidth]{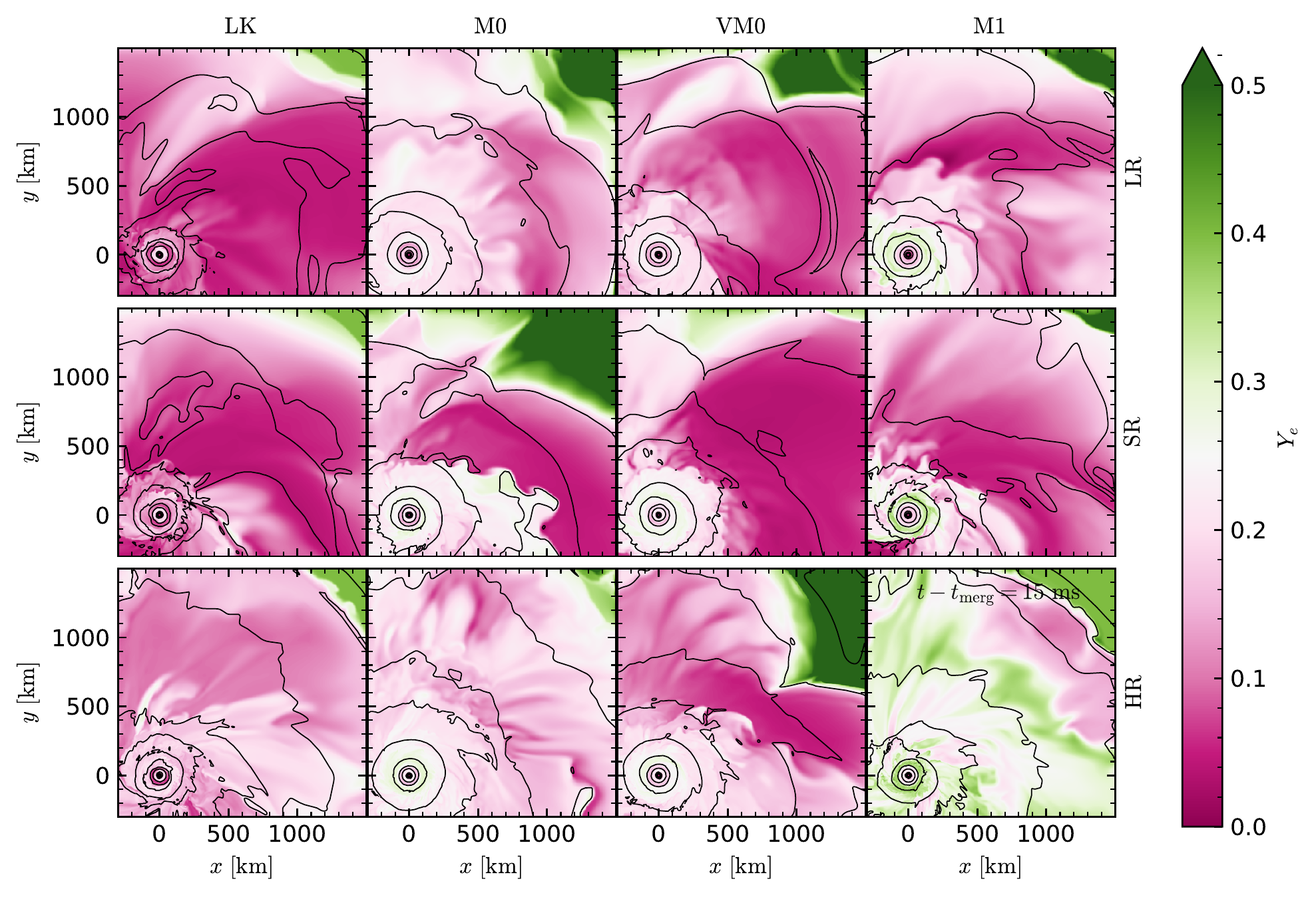}
	\caption{2D snapshot of the $xz$-plane showing the $Y_e$
	         of the material around the remnant \ac{NS} at $\ttmerg = 20$ 
	         ms. The thickest black curve is the isodensity
	         contour $\rho = 10^{13}~\rhocgs$ delimiting 
	         the \ac{NS} remnant, while the others moving 
	         outwards correspond to densities 
	         $\rho = 10^{12},\,10^{11},\,10^{10},\dots~\rhocgs$.
	         Each row correspond to different resolutions, while each
	         column to different micro-physics prescriptions. 
	         Profiles for HY runs 
	         are not reported because neutrinos are not simulated
	         and the electron fraction distribution is frozen at 
	         $Y_e = 0.05$.
	         Note that the profile for M1-HR run is reported at 
	         15 ms \pm{}, close to the last available time.
	         }
	\label{fig:ejecta_2d_xy}
\end{figure*}
\begin{figure*}
	\centering
	\includegraphics[width=0.98\textwidth]{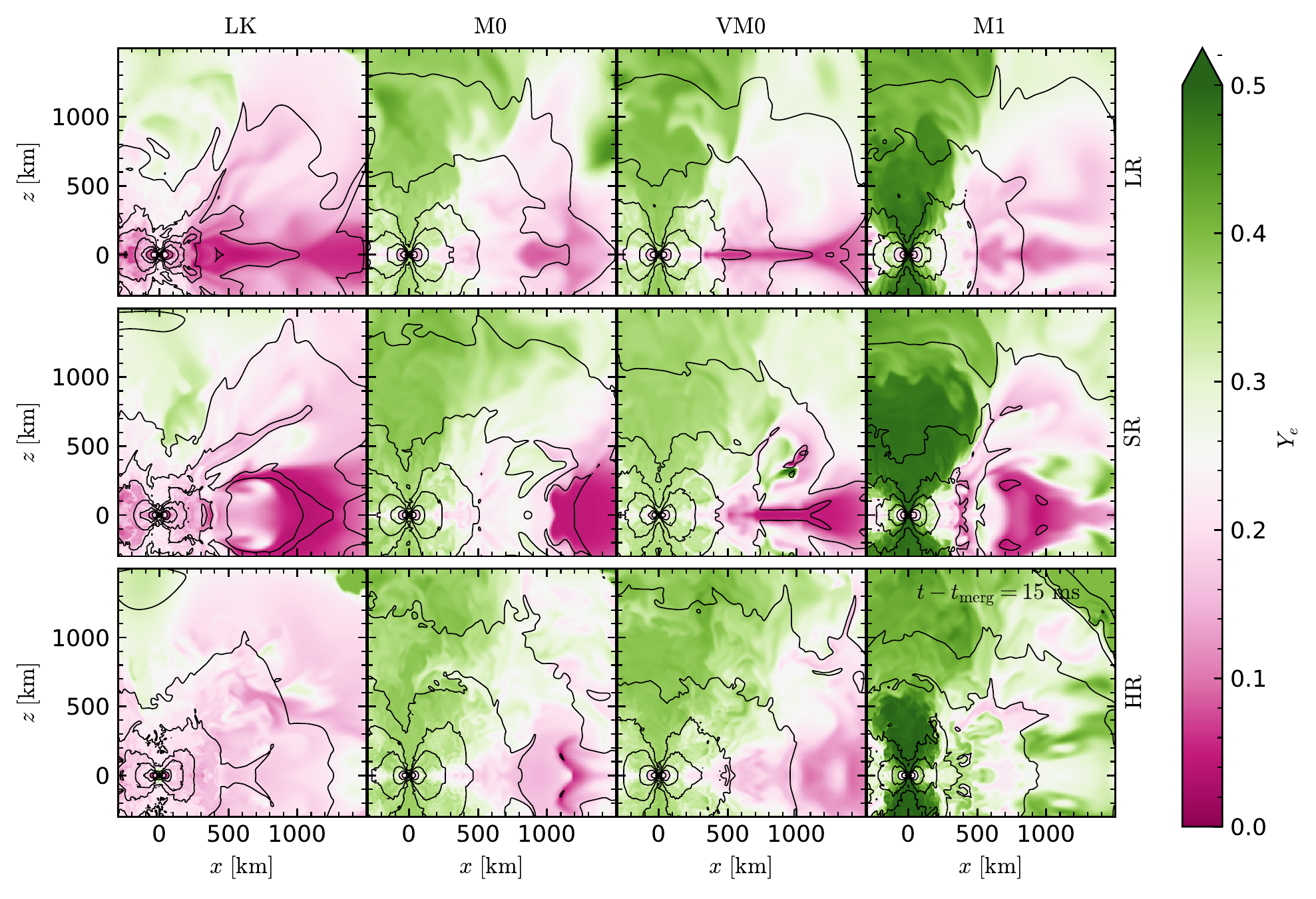}	
	\caption{2D snapshot of the $xz$-plane showing the $Y_e$
	         of the material around the remnant \ac{NS} at $\ttmerg = 20$ 
	         ms. The thickest black curve is the isodensity
	         contour $\rho = 10^{13}~\rhocgs$ delimiting 
	         the \ac{NS} remnant, while the others moving 
	         outwards correspond to densities 
	         $\rho = 10^{12},\,10^{11},\,10^{10},\dots~\rhocgs$.
	         Each row correspond to different resolutions, while each
	         column to different micro-physics prescriptions. 
	         Profiles for HY runs 
	         are not reported because neutrinos are not simulated
	         and the electron fraction distribution is frozen at 
	         $Y_e = 0.05$.
	         Note that the profile for M1-HR run is reported at 
	         15 ms \pm{}, close to the last available time.
	         }
	\label{fig:ejecta_2d_xz}
\end{figure*}

$Y_e$ (first column of Fig.~\ref{fig:ejecta_hist}) exhibits the most
complex behaviours, different among the runs.
To discuss it, we also refer in the following to 
Figs. \ref{fig:ejecta_2d_xy} and \ref{fig:ejecta_2d_xz}, where
we report 2D slices of the $Y_e$ profiles in the $xy$- and $xz$- 
plane respectively.
For HY runs $Y_e$ is frozen at ${\sim}$0.05 because weak
interactions are not simulated and the matter composition does not
change throughout the run with respect to the initial neutrino-less 
weak equilibrium condition. 
For \ac{LK} cases the ejecta mass composition peaks at
$Y_e \approx 0.13-0.17$ (compare also to Tab.~\ref{tab:disk_ejecta}).
No significant fraction of ejecta has $Y_e > 0.35$. 
The material at low 
$Y_e \lesssim 0.15$ is emitted at small latitudes (left-most column of 
Fig.\ref{fig:ejecta_2d_xz}), while for increasing angles $Y_e$
increases, reaching $Y_e \lesssim 0.35$ in the lower-density region
above the remnant \ac{NS}. Matter at high latitudes is 
shock-heated ejecta, therefore hot, 
and is expanding in a region where the disc is not 
present. Under these conditions, the expanding matter becomes
transparent earlier producing electron-positron pairs.
Therefore, positron captures increasing $Y_e$
are more efficient even in absence of neutrino absorption.
For both M0 and VM0 the $Y_e$ distribution gets broader with respect
to \ac{LK}, with a large fraction of matter 
having $Y_e \in [0.2,0.35]$.
This is the effect due to neutrinos radiated by the central object 
and the disc that are absorbed by neutrons in the ejecta, converting
neutrons into protons.
As in the previous case, the low-$Y_e$ material is emitted at lower
latitudes and the $Y_e$ increases for increasing latitudes. 
The peak at $Y_e \approx 0.3$ observed in the left column of 
Fig.~\ref{fig:ejecta_hist} is reached in the high-latitudes, 
low-density ejecta (second and third column of Fig.\ref{fig:ejecta_2d_xz}).
This is because neutrino fluxes are significantly larger at high 
latitudes, due to the presence of the disc at low latitudes.
In M1 runs the trend is similar but even higher values of $Y_e$ are reached.

The histograms in Fig.~\ref{fig:ejecta_hist} show that the peak at 
$Y_e \approx 0.3$ of M0 and VM0 translates to $Y_e \gtrsim 0.425$ when switching to M1. Material
with such a high $Y_e$ is found once again at large latitudes. 
The comparison to M0 runs indicates that accounting for neutrino 
transport with a more complete neutrino scheme provides more 
efficient proton production in the shock-heated ejecta component.
One of the causes of this is that the M0 scheme uses a spherical
grid that assumes neutrinos are only moving radially. On the contrary,
the M1 scheme is solved in the computational grid and the
radiation is evolved according to 3D transport. Neutrinos from
the disc will naturally tend to escape along the 
$z-$ direction, in which the gradient of the optical thickness 
decreases more steeply and the neutrinos mean-free path increases faster,
further irradiating the high-latitude ejecta.

Finite resolution has a clear effect on the ejecta
composition, especially visible at HR.
All runs at LR and SR show a peak at $Y_e \approx 0.05$ that is
due to the tidal component of the ejecta, which is emitted
at early times after merger and maintains the $Y_e$ of the 
two initial stars.
However, for HR runs this component is strongly suppressed for all but 
the run with viscosity. This can be explained by two different 
factors. 
First, the tidal ejecta are expected to be less massive at HR, because
the tidal deformation causing this emission at merger are 
better resolved.
Second, the discs are less massive and geometrically thinner
for HR runs, compared to the others. Therefore, it is easier for neutrinos
to escape from the inner regions and interact with the ejecta, increasing
its $Y_e$. The latter explanation is supported by the fact that 
in VM0-HR run
the disc is not as thin as in the other HR runs and only for this case the 
low-$Y_e$ peak is not heavily suppressed. 
This contributes to explain why
the M1-HR run exhibits such a large $Y_e$ in both the $xy$- and 
$xz$- planes.
On the one hand, the disc is thinner because it is a HR run. 
On the other hand, neutrino fluxes predicted by the M1 scheme
increase the $Y_e$ in the matter more efficiently with respect 
to the M0 scheme.

\section{Nucleosynthesis and Kilonova light curves}
\label{sec:nuc_KN}

\subsection{Nucleosynthesis}

\begin{figure*}
	\centering
	\includegraphics[width=0.98\textwidth]{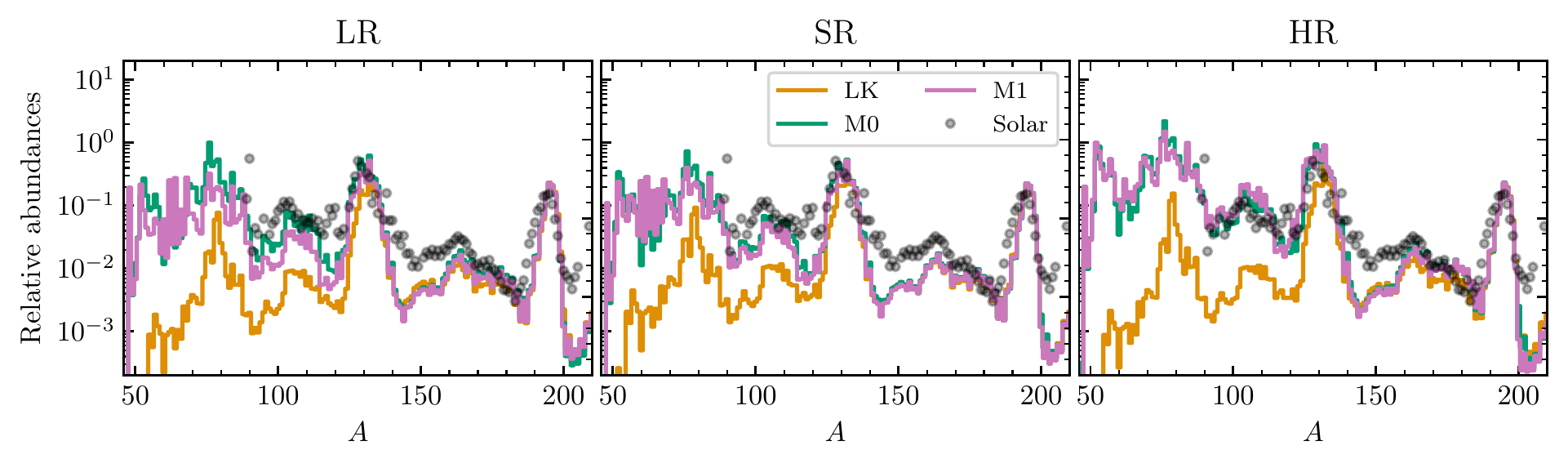}
	\caption{Nucleosynthesis yields comparison between the three
	physical prescriptions against the solar abundancies for 
	the three resolutions
	in each panel. On the $x-$axes the mass number $A$ is reported. 
	The relative abundances on the $y-$axis are expressed in logarithmic
	scale. Consistently with Figs. \ref{fig:ejecta_hist}, the
	analysis is performed up to $\ttmerg \approx 20$ ms for all runs
	except M1-HR, for which $\ttmerg \approx 15$ ms.}
	\label{fig:yields}
\end{figure*}

We compute nucleosynthesis abundances inside the ejecta extracted from our 
simulations according to the procedure described in \citet{Radice:2018pdn}.
The resulting nucleosynthesis yields are shown in Fig.~\ref{fig:nu_lum}. We
compare the results obtained from LK, M0 and M1 runs against the solar 
residual $r$-process abundances from \citet{Arlandini:1999an}.
Abundances are normalised by fixing the overall fraction of elements with $A 
\in [170,\,200]$ to be the same for all set of abundances.
We find that, once the third peak abundaces have been fixed, the abundaces 
predicted by all neutrinos schemes are roughly compatible among them and with 
the solar residual pattern for $A \in [125,\,140]$ (i.e., for the second $r$-
process peak) and $A \in [170,\,200]$. For $A \in [140,\,170]$ and $A>200$ 
yields from all of our simulations significantly differ from the solar 
residuals. Such discrepancies are possibly due to nuclear physics inputs, as 
well as to a lack of suitable physical conditions to efficiently produce 
actinides, see e.g. \citep{Mumpower:2016zye,Wu:2022wqm}.
The \ac{LK} runs heavily underestimate the abundances for $A<120$. This is a 
direct consequence of the fact that $Y_e$ is lower in 
the ejecta for these cases.
In the runs with M0 and M1 the abundances for $A<120$ are closer among them 
and to the solar residuals, compared to LK.

Increasing the resolution does not change the abundances in runs 
with LK, which also at high resolution significantly differ from 
the solar residual abundances for $A<120$.
For M0 and especially for M1 runs the predictions at 
HR better match the solar abundances 
for the entire range of nuclear masses (still with the exceptions discussed above).

Our results confirm the relevant role of neutrinos emission and absorption in shaping the nucleosynthesis yields from the early time ejecta of BNS mergers \citep[see, e.g.,][]{Wanajo:2014wha,Goriely:2015fqa,Martin:2017dhc,Radice:2018pdn}. Abundances obtained in our HR simulations employing the M0 or M1 schemes are compatible among them and reproduce well the observed solar residual pattern. However, models featuring neutrino cooling alone underestimate the abundances of light $r$-process elements, since neutrino reabsorption is required to produce the ejecta conditions suitable for the production of those elements.

\subsection{Kilonova light curves}

We compute synthetic kilonova light curves following the approach outlined 
in \cite{Wu:2021ibi} and using the SNEC radiation-hydrodynamics
Lagrangian code \citep{Morozova:2015bla}. 
Accordingly, the dynamical ejecta computed from our simulations are further 
evolved with SNEC up to 15 days \pm{}. 
The corresponding light curves are presented
using the AB magnitude system
\begin{equation}
m_\mathrm{AB} = -2.5 \log_{10} \left(\frac{\int f_\nu(h\nu)^{-1} e(\nu) d\nu}
{\int{3631 \mathrm{Jy}(h \nu)^{-1} e(\nu) d\nu}}\right)
\end{equation}
where here $\nu$ is the light frequency, $f_\nu$ is the observed flux 
density at frequency $\nu$ from a distance of $40$ Mpc and $e(\nu)$
are filter functions for different Gemini bands. We refer to 
\cite{Wu:2021ibi} for more details.

\begin{figure*}
	\centering
	\includegraphics[width=0.95\textwidth]{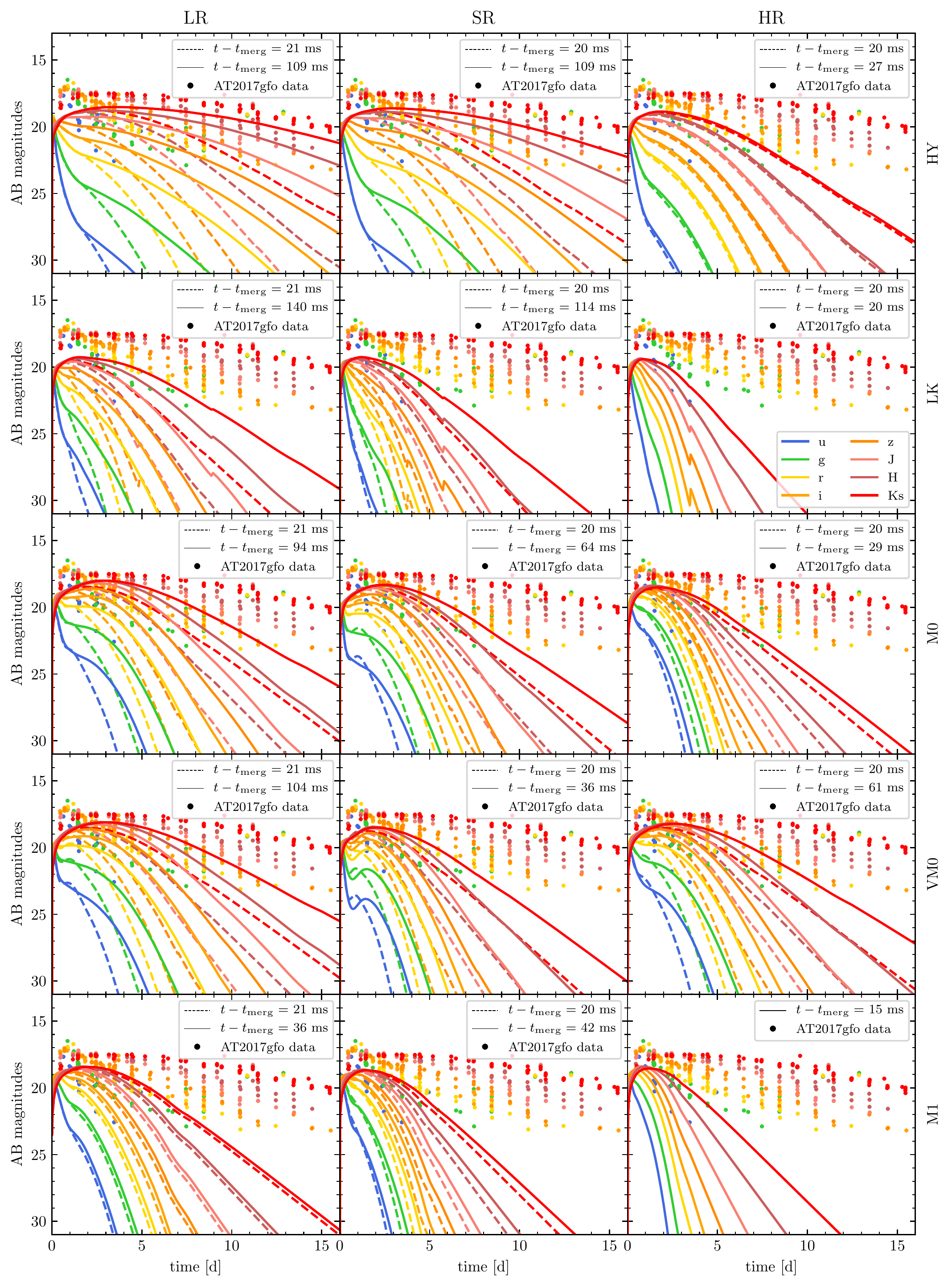}
	\caption{Light curves calculated with SNEC. We report the
	AB magnitudes as a function of days after merger. The light curves
	corresponds to several Gemini bands and are calculated from the ejecta
	extracted at $R = 443$ km from the system at a common time $20$ ms
	\pm{} (dashed lines) and at the end time of each simulation 
	(solid lines). Dots correspond to 
	the data of the kilonova event AT2017gfo for the same bands.}
	\label{fig:KNEC_light_curves}
\end{figure*}

In Fig. \ref{fig:KNEC_light_curves}  we compare the AB magnitudes 
at different bands to the
electromagnetic transient $\mathrm{AT2017gfo}$ associated to the
\ac{BNS} merger event GW170817 \citep{Villar:2017wcc}.
As input for the SNEC code, we consider the ejecta extracted at two different times: at 20 ms \pm{} (dashed lines) and at the end of the simulation (solid lines), with the exception of the HR-M1 run. Clearly, the different simulation lengths impact on the light curve due to the different ejecta masses, but also due to the composition.
AT2017gfo is significantly brighter than any of our light curves. Nonetheless the hierarchy of the colors 
is correct at $\sim$4 days, whereas the 1 day emission has 
a blue peak that cannot be explained with dynamical 
ejecta we are considering here.
The fact that our analysis does not reproduce the
data is expected for many reasons. 
First, the \ac{BNS} we simulate is not targeted to the event GW170817; 
in particular it has lower mass and symmetric mass ratio, which implies
smaller ejecta masses and therefore dimmer light curves. 
Second, our simulations are too short and cannot 
capture the full evolution of the \pm{} disc. Therefore, 
ejecta emitted at secular timescales (seconds after merger) 
is missing. Crude estimates of later outflows emission can be
made by extrapolating in time~\citep{Wu:2021ibi}, 
but we do not attempt this here.
Third, multidimensional effects and viewing angle
can have a strong impact on the kilonova emission 
\citep[e.g.,][]{Perego:2017wtu,Kawaguchi:2019nju,Korobkin:2020spe}
but are neglected here. For AT2017gfo, spherically symmetric kilonova
models are ruled out with high confidence
\citep{Villar:2017wcc,Perego:2017wtu,Breschi:2021tbm}.
In the following we focus on the differences 
seen for different microphysics.

For HY runs we obtain that light curves corresponding to the
$K_s$, $H$ and $J$ band are only a few magnitude larger than AT2017gfo data, 
especially when we consider ejecta production at ${\sim} 109$ ms
\pm{} (solid lines in the LR and SR cases). By contrast, dynamical ejecta 
alone produce significantly dimmer light curves (dashed lines), 
in particular at late time after the peaks.
Despite the usually long simulation lengths, for \ac{LK} runs the ejecta mass 
is smaller and this produces dimmer light curves compared to HY runs, 
considering both the early ejecta and those at the end of the simulations.
The jumps that we observe in these curves are an artefact of the SNEC code.
For M0, VM0 and M1 we obtain brighter light curves at all bands with respect to LK, as a consequence of the fact that more ejecta mass, characterised by a larger $Y_e$, is produced.
When considering only the early ejecta (dashed lines), M0, VM0 and M1 produce very compatible light curves, due to the very similar ejecta properties, see Sec. \ref{sec:ejecta} and Table (\ref{tab:disk_ejecta}). M1 light curves are slightly dimmer due to the faster and less opaque ejecta, which translate in a faster kilonova evolution after the peaks. Differences become more pronounced when light curves are computed using the ejecta at the end of the simulations, since M1 runs were evolved for shorter post-merger times and produced systematically less ejecta mass.

Finite resolution does not significantly impact the light curves.
Our analysis shows that the light curves are very sensitive both to the inclusion of neutrino reabsorption in optically thin conditions and to the cumulative
time during which ejecta are measured. During this time not only the
ejecta mass, but also the ejecta composition changes due to the different
emission mechanisms at different timescales. The better accuracy provided by the M1 scheme with respect to the M0 one seems to have a minor impact on the kilonova light curves due to the good agreement in the ejecta properites between the two schemes, when the simulations have comparable lengths. Future simulations 
will extend these results by also considering the winds from the 
viscous \pm{} phase and taking into account non-spherical geometries.

\section{Neutrino luminosity}\label{sec:nu_luminosity}

\begin{figure*}
	\centering
	\includegraphics[width=0.98\textwidth]{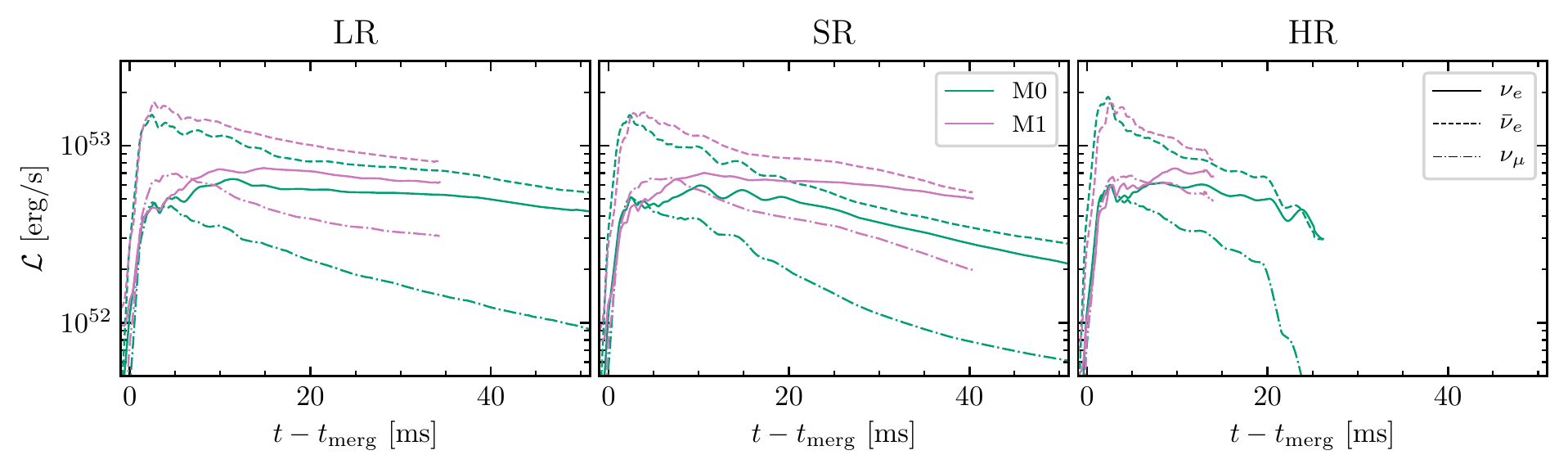}
	\caption{Neutrino luminosity comparison between M0 and M1
	simulations for all three neutrino species and for the 
	three resolutions in each panel. 
	Neutrinos are extracted at a radius $R_\mathrm{M0} = 756$ km
	for the M0 case and $R_\mathrm{M1} = 443$ km for the M1 case.
	The data is smoothed using a rolling average with width 1 ms.
	Time is shifted by the time of merger
	and by the time of flight of neutrinos to the corresponding detector. 
	The luminosity is expressed in cgs units and in 
	logarithmic scale. 
	}
	\label{fig:nu_lum}
\end{figure*}

\begin{figure*}
	\centering
	\includegraphics[width=0.98\textwidth]{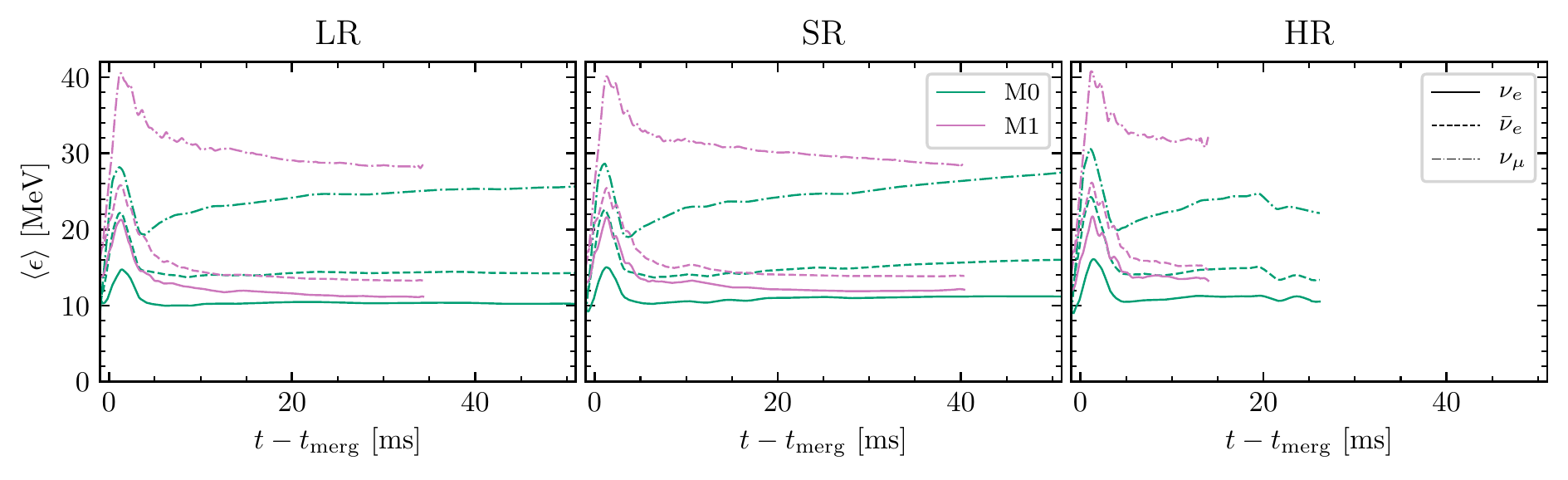}
	\caption{Neutrino average energy comparison between M0 and M1
	simulations for all three neutrino species and for the 
	three resolutions in each panel. 
	Neutrinos are extracted at a radius $R_\mathrm{M0} = 756$ km
	for the M0 case and $R_\mathrm{M1} = 443$ km for the M1 case.
	The data is smoothed using a rolling average with width 1 ms.
	Time is shifted by the time of merger
	and by the time of flight of neutrinos to the 
	corresponding detector.}
	\label{fig:nu_energy}
\end{figure*}
We now discuss the impact of different microphysics and finite 
resolution effects on the neutrino emission in our simulations.
In Fig.\ref{fig:nu_lum} we show
the angle integrated neutrino luminosity for the three neutrino
species we simulate, comparing M0 and M1 neutrino schemes 
for every resolution. Hereafter, we consider one representative
heavy flavour neutrino species denoted as $\nu_\mu$ with properties
calculated as averages over the four neutrino species constituting 
$\nu_x$.
Neutrino luminosities for every species present a peak immediately after 
merger at $\L \approx 10^{52}-10^{53}$ erg~s$^{-1}$. 
The hierarchy $\L_{\nu_\mu} < \L_{\nu_e} < \L_{\bar{\nu}_e}$
that we observe in the neutrino luminosity evolution is consistent with previous results \citep[see, e.g.,][]{Ruffert:1996by,Rosswog:2003tn,Sekiguchi:2015dma,Foucart:2016rxm,Cusinato:2021zin}
and it is explained as follows.
Electron antineutrinos are the most abundant species because 
the positron captures
on free neutrons are favoured in the neutron rich ($Y_e \approx 0.1$) 
matter with temperatures of tens of MeV. 
Electron neutrinos are produced instead mostly due
to capture of electrons on protons, which are however not favoured
due to the initially low proton abundance. Heavy flavour 
neutrinos are produced by matter with temperature of tens of MeV 
emitted from the bouncing remnant. The reactions producing 
heavy flavour neutrinos are electron-positron annihilation and 
plasmon decay which are highly dependent on temperature. 
As the remnant stabilises and cools down, production of 
heavy flavour neutrinos lowers, while electron/positron
captures keep happening in the highest density region of 
the accretion discs producing electron neutrinos and antineutrinos.

Comparing M0 and M1 runs, we observe that the same luminosity 
hierarchy is maintained, but M1 scheme predicts larger
neutrino brightnesses. The largest difference is observed for 
heavy flavour neutrinos, where neutrinos in M1 runs are ${\sim} 50\%$ 
brighter than neutrinos in M0 runs.

This is in contrast with \cite{Radice:2021jtw},
where neutrino luminosities for all species were found a factor 
${\sim}3$ larger for M0 compared to M1 scheme.
The reason for this difference is that here the neutrino luminosities
are computed integrating the proper radiation flux calculated as 
$\mathcal{F}^i = \sqrt{\gamma} (\alpha F^i - \beta^i E)$ over the sphere,
where $E$ and $F^i$ are the radiation energy density and the radiation
flux in the Eulerian frame, respectively.
By contrast, in \cite{Radice:2021jtw} neutrino luminosities were 
computed from the covariant expression $\mathcal{F}_i = \sqrt{\gamma} F_i$,
which is an approximation valid for large extraction radii. 
Phisically, both quantities are expected to have the same asymptotic value
${\sim}E$, but we find that at our finite radius they differ by a 
factor ${\sim}3$. 

When a \ac{BH} forms, the emission of heavy flavour neutrinos
abruptly stops because this component is emitted from the remnant \ac{NS}.
As resolution increases, we observe an increase in the 
luminosities at early times, within 15 ms post merger. 
This is largely explained by the 
fact that thinner discs are formed at this resolution, which allow 
neutrinos to diffuse more easily and with shorter timescales.
Larger electron antineutrino luminosities at HR,
for both M0 and M1 schemes, are in agreement with the fact that larger
electron fractions are found in the ejecta distributions for HR.

In Fig.\ref{fig:nu_energy} we report the neutrino average energies for 
the same runs. For both M0 and M1 schemes, the energies
peak at $2-3$ ms \pm{} before reaching a quasi-steady evolution at later times, and follow a hierarchy 
$\langle\epsilon_{\nu_e}\rangle < \langle\epsilon_{\bar{\nu}_e}\rangle < 
\langle\epsilon_{\nu_\mu}\rangle$. 
Focusing on M1 runs, neutrino average energies for heavy
lepton neutrinos peak at $40$ MeV and then 
it decreases below ${\sim}30$ MeV within few tens of ms.
Electron antineutrinos and neutrinos follow a similar behaviour 
also with similar timescales,
reaching their maxima at ${\sim}25$ MeV and ${\sim}20$ MeV, and decreasing
to ${\sim}12$ MeV and ${\sim}10$ MeV, respectively.
Runs with M0 scheme systematically underestimate the energies in the 
first $\ttmerg = 20$ ms by $\approx 30 \%$ with respect to M1 runs. 
We also note that the average energy of the heavy lepton neutrinos 
increases with time, reaching values comparable to the ones simulated
with M1 scheme, within tens of ms \pm{}.

The quantitative differences in the $\nu_e$ and $\bar{\nu}_e$ energies between the two sets of 
runs possibly originate from different causes. M1 simulations tend to produce more massive and inflated discs. Such discs have more extended neutrino surfaces characterised by lower decoupling temperatures. At the same time, the \MO scheme properly models the diffusion of
neutrinos inside the remnant up to the emission at the
neutrino surface and their thermalization~\citep{Radice:2021jtw}.
In M0 schemes, instead, the diffusion rate is estimated based on local properties and thermalization effects of diffusing neutrinos are not taken into account.
For heavy flavour neutrinos the situation is opposite: neutrinos decouple from matter deep inside the remnant, further diffusing through quasi-isothermal scattering inside the disc. While an M1 scheme is able to catch this effect, retaining larger $\nu_{\mu}$ mean energies, the M0 computes the luminosities and mean energies considering neutrinos in equilibrium with matter everywhere inside the last scattering surface, providing at the same time lower mean energies and larger luminosities. With time, the disc becomes more compact and the diffusion atmosphere reduces in size, so that $\nu_{\mu}$ mean energies become comparable.

The average neutrino energies are not largely influenced by resolution
effects. This is expected because the neutrinospheres are mostly 
determined by the density profile inside the 
disc \citep{Endrizzi:2019trv}, which we showed to be robust with
resolution.

Finally, it is interesting to notice that the value of the high electron fraction peak in the $Y_e$ distribution in the ejecta extracted from the M1 runs (and observed at high latitudes) is close to the equilibrium electron fraction, $Y_{e,{\rm eq}}$. The latter can be estimated using eq. 77 of \citet{Qian:1996xt}. Assuming, according to our neutrino luminosities and mean energies around 10~ms \pm, $\mathcal{L}_{\bar{\nu}_e} \approx 3/2~\mathcal{L}_{\nu_e}$, $\langle \epsilon_{\nu_e} \rangle \approx 12~{\rm MeV}$ and $\langle \epsilon_{\bar{\nu}_e} \rangle \approx 14~{\rm MeV}$, 
we find $Y_{e,{\rm eq}} \approx 0.46$. This means that in the region above the massive \ac{NS} absorption rates in the M1 runs are high enough to approach weak equilibrium, and the differences with the M0 results are mostly due to the rates values, rather than to differences in the relative luminosities or mean energies.

\section{Conclusions}\label{sec:conclusion}

In this work we performed the first systematic study of the impact
of different treatments of neutrino transport on the computation of
multi-messenger observables from a \ac{BNS} merger. Our work is based on
ab-initio 3+1 \ac{NR} simulations  performed at 
three resolutions for each microphysics prescription and up to
resolutions of ${\sim}123$~meters (HR) in the strong-field region. 
We simulated and compared pure hydrodynamics (HY), leakage (\ac{LK}), leakage+M0 (M0)
and M1 neutrino transport schemes. The M0 series of simulations was
also repeated with the \ac{GRLES} subgrid scheme for MHD turbulent
viscosity. The simulations considered a BNS merger forming a short-lived
remnant; they cover the GW-dominated \pm{} phase and last at least ${\sim}15$~ms and up to 140 ms \pm{}.

Our analysis indicates that the gravitational collapse of the short
lived remnant is mainly determined by the emission of GWs and angular
momentum transport. Turbulent viscosity can significantly affect the
collapse by stabilizing the remnant, whereas the impact of different
neutrino schemes is negligible. \ac{BH} collapse happens as the
remnant approaches the maximum density of the corresponding cold,
$\beta$-equilibrated spherically symmetric equilibria 
\citep{Perego:2021mkd},
in particular for $\rhomax \gtrsim 70\% \rho_\mathrm{max}^\mathrm{TOV}$.
The remnant's stability and the time of collapse are strongly affected
by the grid resolutions. In our setup, high resolutions generically
induce an earlier collapse while numerical effects at low
resolutions can stabilise the remnant. 
Nonetheless, we find that the remnant's bulk dynamics can be robustly 
studied using the gauge-invariant curves of binding energy and maximum 
rest-mass density. As shown in Fig.~\ref{fig:rho_ebind}, these 
quantities are strongly correlated and the correlation is not 
sensitively dependent on the grid resolution.
This implies the possibility of probing the maximum remnant
densities from inferences of the emitted GW energy \citep{Radice:2016rys}. 

Accretion discs of initial masses up to $\Discmass \approx 0.2~\Msun$
form around the remnant \ac{NS} during merger. The disc masses depend on
the microphysics prescription used: \ac{LK} simulations produce the
least massive disc, while HY simulations produce the most
massive disc (for sufficiently high resolutions).
In general, including neutrino transport leads to more inflated discs
with respect to pure hydro. 
The electron fraction of disc matter at low latitudes 
reaches values of $\approx 0.25$ for \ac{LK} schemes, but is
larger in M0 and VM0 runs comparing matter shells at same density. At lower
densities or higher latitudes, M0 schemes predicts $Y_e \gtrsim 0.3$. 
The M1 scheme leads to the largest $Y_e \gtrsim 0.42$.
Increasing the grid resolution leads to the formation of less massive,
more compact discs but it does not significantly affect their composition.
However, the disc mass and accretion rates are heavily dependent on black formation, which in turn is
affected by resolution (see above). 
Overall, our analysis indicates that advanced transport scheme are absolutely
necessary in future long-term disc evolutions, and \ac{LK} schemes should be
abandoned. At the same time, \pm{} simulations at mesh
resolutions above 200~meters seem insufficient to deliver 
quantitative results for astrophysical predictions.

Simulations with M1 transport show the emergence of a neutrino trapped gas
in the remnant's \ac{NS} core \citep{Foucart:2016rxm,Perego:2019adq,Radice:2021jtw}.
The neutrino gas locally decreases the temperature and increases $Y_e$ by
${\sim}30~\%$ comparing to \ac{LK} runs. We do not observe changes in 
the pressure and consequent alterations in the gravitational collapse 
in our models.
The abundances of the neutrino species in the trapped gas
are in the hierarchy $Y_{\nu_e} < Y_{\nu_x} < Y_{\bar{\nu}_e}$, 
that can be understood from the thermodynamics conditions in the 
remnant \ac{NS} core \citep{Perego:2019adq}. 

\ac{GW} emission is not significantly affected by
microphysics in the considered BNS, despite the latter being a
binary that produces a short-lived remnant close to the collapse.
The main GW properties can be robustly extracted from simulations with
at least SR resolution. Our \pm{} faithfulness analysis indicates that,
at SR and HR resolutions, the waveform quality is sufficient
for an accurate modeling of \pm{} signals. This precision is sufficient for both detecting \pm{} signals with matched-filter
analyses and for constraining the \ac{EoS} 
at extreme matter densities with third generation observations \citep{Breschi:2022xnc}. 
In contrast to Refs.~\cite{Most:2022yhe,Hammond:2022uua}, 
our high-resolution M1 simulations do not show any 
evidence of a significant out-of-equilibrium and bulk viscosity 
effects in the GWs. The key
differences between our work and previous ones is the consistent treatment
of neutrino radiation and the higher grid resolution (more than a factor
3 higher in our HR runs). 

In our simulations, ejecta of $\mej \gtrsim 2\times 10^{-3}$ are launched
during merger, with the smallest (largest) ejection measured in \ac{LK} (HY) runs.
Increasing the resolution typically decreases the ejecta
mass. The largest deviation is $\approx 50\%$ (VM0-SR and VM0-HR),
that could be taken as an estimate of the current \ac{NR} uncertainties.
We find that the early ejecta mass as a function  
of the ejecta velocity can be modelled with a two-components broken
power-law of type 
$\propto (\beta\gamma/\left(\beta\gamma\right)_{\beta_0})^{-s}$,
with $\beta_0 \in [0.3,\,0.45]$. 
The most massive and slower component has
$s_\mathrm{KN}\approx0.64-1.6$ for $v < \beta_0c$, whereas the fast tail 
component has a steeper profile,
$s_\mathrm{ft}\gtrsim 4-11$, and masses $\mej(v_\infty > \beta_0 c) \approx 10^{-5}-10^{-4}~\Msun$. 
The $Y_e$ distribution in the dynamical ejecta largely depends on the
simulated microphysics. In all LR and SR runs we observe very neutron rich 
ejecta component at low latitudes, corresponding to the tidal component.
The LK scheme predicts a second peak in the $Y_e$ distribution
at $Y_e\approx 0.13-0.17$, corresponding to the shock-heated component. 
Matter leptonization due to neutrinos emitted by the 
remnant and reabsorbed in the ejecta produces a broader peak in the $Y_e$ distribution
of the runs with M0, spanning $Y_e \approx 0.2-0.35$.
M1 simulations show an additional high-$Y_e$ peak at $Y_e \approx 0.425$, 
corresponding to material emitted at high latitudes.
With the only exception of VM0-HR, we note that the low-$Y_e$ peak is 
suppressed in HR runs due to two reasons.
First, the tidal component is smaller because the remnant star is 
more compact than in lower resolution runs. 
Second, discs at HR are thinner and less massive than
lower resolutions ones, thus less opaque to neutrinos. This effect is mitigated in VM0-HR run, where viscosity effects
produce a larger disc compared to the other HR runs.

Our results confirm the relevant role of neutrinos emission and absorption in shaping the nucleosynthesis yields from the early time ejecta. Both M0 or M1 schemes deliver, at high-resolutions, comparable abundances\footnote{
Note both ejecta have a component with $Y_e \gtrsim 0.2-0.35$.}
and  reproduce well the observed solar residual pattern. On the contrary, the \ac{LK} scheme alone underestimates the abundances of light $r$-process elements, since neutrino reabsorption is required to produce the ejecta conditions suitable for the production of those elements.  

The calculated kilonova light curves are rather robust against grid resolution but are very sensitive to the cumulative time during which ejecta are measured and to the effect of neutrino irradiation. 
Larger ejecta masses and lower $Y_e$ generate brigther kilonova light
curves. Accordingly, HY (\ac{LK}) runs produce the brightest (dimmest) kilonovae as shown in Fig.~\ref{fig:KNEC_light_curves}. However, the largest light curve variations in the plot are associated to the use of the ejecta calculated over different time intervals. During these times the ejecta mass increases and the ejecta composition changes due to an early wind component summing up to the dynamical ejecta. 
These results highlight, once again, the critical need for long-term merger and \pm{} simulations with realistic microphysics for the reliable prediction of the EM counterparts to mergers.

We find neutrino luminosities of the order of 
$\L \approx 10^{52}-10^{53}$. 
M0 runs slightly underestimate the neutrino luminosity
with respect to M1 runs, for each simulated neutrino species. This
difference is largest in the case of heavy flavour neutrinos, because
only M1 schemes are able to simulate their diffusion inside the disc.
However, in both cases the two schemes consistently predict the hierarchy
$\L_{\nu_\mu} < \L_{\nu_e} < \L_{\bar{\nu}_e}$.
These results confirm previous findings 
\citep{Foucart:2016rxm,Radice:2021jtw} and
stresses the importance of using M1 schemes for detailed predictions.
Larger $\L_{\bar{\nu}_e}$ are found at HR, which is explained
by the presence of thinner discs. This is consistent with the ejecta composition
summarised above and in particular it is related to the suppression of the
low-$Y_e$ peak in HR $Y_e$ distribution. 

Our work highlights the fact that both resolution and microphysics
can have a significant impact on 
the observables predicted by a BNS merger simulation.
Future work will be focused on extending M1 simulations to different
binaries and for longer times after merger. 
Our results indicate that advanced neutrino schemes, like the M1, and
sub-grid-MHD effects are likely necessary physics input for an
accurate prediction of the winds from the remnant. At the same time,
high-resolution simulations appear essential for robust results in
long-term evolutions.

\section*{Acknowledgements}
F.~Z. and S.~B. acknowledge support by the EU H2020 under ERC Starting
Grant, no.~BinGraSp-714626.
SB acknowledges support from the Deutsche Forschungsgemeinschaft, DFG,
project MEMI number BE 6301/2-1. 
D.~R. acknowledges funding from the U.S. Department of Energy, Office
of Science, Division of Nuclear Physics under Award Number(s)
DE-SC0021177 and from the National Science Foundation under Grants
No. PHY-2011725, PHY-2020275, PHY-2116686, and AST-2108467.
A. P. acknowledges support from the INFN through the TEONGRAV initiative and thanks the Theoretisch-Physikalisches Institut at the Friedrich-Schiller-Universit{\"a}t Jena for its ospitality.
\ac{NR} simulations were performed at the
ARA cluster at Friedrich Schiller University Jena, 
SuperMUC\_NG at the Leibniz-Rechenzentrum (LRZ) Munich and
HPE Apollo Hawk at the High Performance Computing Center Stuttgart (HLRS).
The ARA cluster is funded in part by DFG grants INST
275/334-1 FUGG and INST 275/363-1 FUGG, and ERC Starting Grant, grant
agreement no. BinGraSp-714626.
The authors acknowledge the Gauss Centre for Supercomputing
e.V. (\url{www.gauss-centre.eu}) for funding this project by providing
computing time on the GCS Supercomputer SuperMUC-NG at LRZ
(allocation {\tt pn68wi}).
The authors acknowledge HLRS for funding this project by providing
access to the supercomputer HPE Apollo Hawk under the grant
number {\tt INTRHYGUE/44215}.
The authors acknowledge XSEDE for funding this project by providing
access to the supercomputers Bridges2 and Expanse under the allocation
{\tt TG-PHY160025}.
This research used resources of the National Energy Research
Scientific Computing Center, a DOE Office of Science User Facility
supported by the Office of Science of the U.S.~Department of Energy
under Contract No.~DE-AC02-05CH11231. Computations for this research
were also performed on the Pennsylvania State University's Institute for
Computational and Data Sciences' Roar supercomputer.

\section*{Data Availability}
 
Data generated for this study will be made available upon 
reasonable request to the corresponding authors.


\bibliographystyle{mnras}

\bsp	
\label{lastpage}
\end{document}